\documentclass[aps,prb,reprint,superscriptaddress,amsmath]{revtex4-2}
\pdfoutput=1
\usepackage{graphicx}
\usepackage{enumerate}
\usepackage{refcount}
\usepackage{fmtcount}
\usepackage{amssymb}
\usepackage{dsfont}
\usepackage{hyperref}
\usepackage[usenames,dvipsnames]{color}
\usepackage{circledsteps}
\usepackage{enumitem}
\usepackage{pdfpages}

\makeatletter
\AtBeginDocument{\let\LS@rot\@undefined}
\makeatother

\hypersetup{pdfstartview=FitH,pdfpagemode=UseOutlines,colorlinks,
citecolor=blue,linkcolor=blue,urlcolor=blue}

\graphicspath{{.}{./Anisotropic_figures/}}

\newcommand{\s}{\sigma}

\newcommand{\dg}{\dagger}
\newcommand{\sdn}{\downarrow}
\newcommand{\su}{\uparrow}

\definecolor{blue(munsell)}{rgb}{0.0, 0.5, 0.69}

\definecolor{aaron}{rgb}{0.6, 0.6, 0.8}

\definecolor{joh}{rgb}{0.25, 0.6, 0.8}

\begin{document}

\title{Phase diagram of the anisotropic triangular lattice Hubbard model}

\author{Aaron Szasz}
	\email[]{aszasz@perimeterinstitute.ca}
	\affiliation{Perimeter Institute for Theoretical Physics, Waterloo, Ontario N2L 2Y5, Canada}
	\affiliation{Materials Sciences Division, Lawrence Berkeley National Laboratory, Berkeley, California 94720, USA}
\author{Johannes Motruk}
	\affiliation{Department of Physics, University of California, Berkeley, California 94720, USA}
	\affiliation{Materials Sciences Division, Lawrence Berkeley National Laboratory, Berkeley, California 94720, USA}

\date{\today}

\begin{abstract}
In a recent study [Phys. Rev. X \textbf{10}, 021042 (2020)], we showed using large-scale density matrix renormalization group (DMRG) simulations on infinite cylinders that the triangular lattice Hubbard model has a chiral spin liquid phase.  In this work, we introduce hopping anisotropy in the model, making one of the three distinct bonds on the lattice stronger or weaker compared with the other two.  We implement the anisotropy in two inequivalent ways, one which respects the mirror symmetry of the cylinder and one which breaks this symmetry.  In the full range of anisotropy, from the square lattice to weakly coupled one-dimensional chains, we find a variety of phases.  Near the isotropic limit we find the three phases identified in our previous work: metal, chiral spin liquid, and 120$^\circ$ spiral order; we note that a recent paper suggests the apparently metallic phase may actually be a Luther-Emery liquid, which would also be in agreement with our results.  When one bond is weakened by a relatively small amount, the ground state quickly becomes the square lattice N\'{e}el order.  When one bond is strengthened, the story is much less clear, with the phases that we find depending on the orientation of the anisotropy and on the cylinder circumference.  While our work is to our knowledge the first DMRG study of the anisotropic triangular lattice Hubbard model, the overall phase diagram we find is broadly consistent with that found previously using other methods, such as variational Monte Carlo and dynamical mean field theory.
\end{abstract}

\maketitle


\section{Introduction}

Over the past few decades, both theoretical and experimental works have pointed to the existence of quantum spin liquids, states for which spin degrees of freedom remain disordered down to zero temperature~\cite{Balents2010,Savary2017,Zhou2017}.  The study of spin liquids has been especially intensive since the experimental identification in 2003~\cite{Shimizu2003} of the organic crystal $\kappa$-(BEDT-TTF)$_2$Cu$_2$(CN)$_3$, abbreviated as $\kappa$-Cu, as a candidate material. The spins in this compound are arranged on a triangular lattice in two-dimensional layers and exhibit no sign of ordering down to temperatures that are several orders of magnitude lower than the spin coupling constant.  In the intervening years, many other triangular lattice materials have been experimentally demonstrated to have a lack of magnetic order down to extremely low temperatures~\cite{Itou2008,Itou2009,Li2015,Shen2016,Law2017,Ribak2017,Shen2017,Zeng2019published,Li2019published,Sarkar2020}. Meanwhile, theoretical work has identified a wide variety of different spin liquid states that might be realized in these systems, ranging from gapped topological states to gapless states with or without a spinon Fermi surface~\cite{Lee2005,Motrunich2005,Lee2007,Qi2009,Grover2010,Mishmash2013}.
Attempts to explain observations in spin liquid candidate materials have recently also focused on the role of disorder~\cite{Furukawa2015,Saito2018,Kimchi2018,Kimchi2018a,Lu2018,Wu2019,Knolle2019,Riedl2019,Choi2019,Dey2020,Volkov2020,Pustogow2020,Miksch2020,Baek2020}.

The correspondence between the predictions from theory and the measured behavior in experiments remains muddled, in part because in some cases there is controversy even about the properties of the materials. In the case of $\kappa$-Cu, specific heat measurements~\cite{Yamashita2008} corroborate the presence of gapless excitations, while thermal conductivity measurements~\cite{Yamashita2008b} and a recent electron spin resonance study~\cite{Miksch2020} suggest a spin-gapped ground state. In another triangular lattice material that has attracted significant interest, EtMe$_3$Sb[Pd(dmit)$_2$]$_2$, it remains under debate whether thermal conductivity measurements demonstrate the existence of mobile gapless excitations~\cite{Yamashita2010,Ni2019,BourgeoisHope2019,Yamashita2019}.
At the same time, there are many different theoretical predictions, which could potentially explain all or parts of the measured behavior, but none of which have been conclusively demonstrated to be correct.

In light of this profusion of possibilities, it is essential to make sure that the models we study do indeed match the actual materials.  
While many spin liquid candidates are believed to be described by the Hubbard model on a nearly perfect triangular lattice, there are typically measured anisotropies on the order of 10 or 20\%, although the precise values are still under debate, in particular for the case of $\kappa$-Cu~\cite{Komatsu1996,Nakamura2009,Kandpal2009,Koretsune2014}.  Additionally, spin liquid-like behavior has been observed in Cs$_2$CuCl$_4$~\cite{Coldea2001}, which is described by a highly anisotropic triangular lattice; the importance of the anisotropy is evident from the fact that more general compounds of the form Cs$_2$CuCl$_{4-x}$Br$_x$, which differ from Cs$_2$CuCl$_4$ in the degree of anisotropy~\cite{Foyevtsova2011,Thallapaka2019}, show a variety of magnetic orders~\cite{Cong2011,Tutsch2019}.

\begin{figure}
\includegraphics[width = 0.48\textwidth]{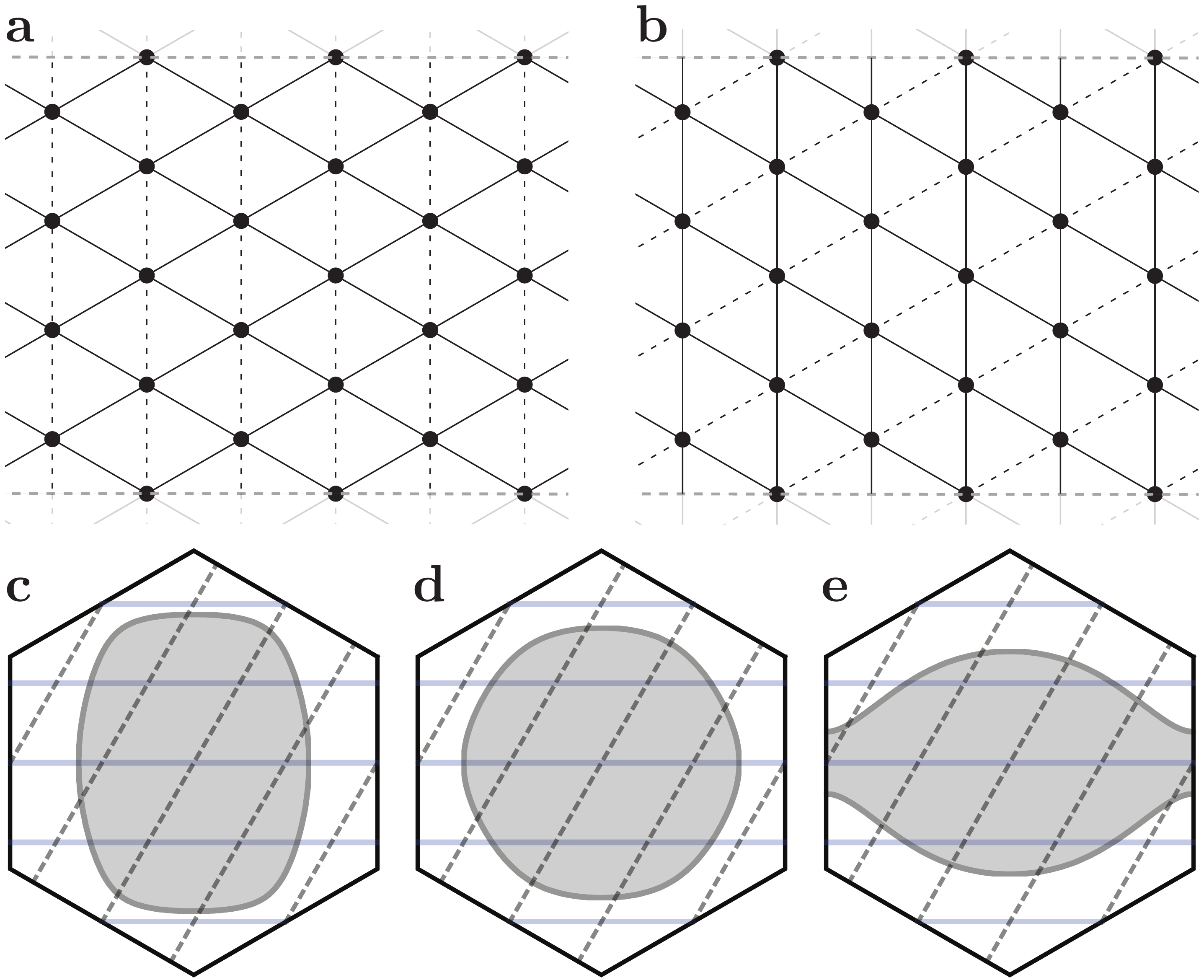}
\caption{{\bf (a)}-{\bf (b)} Triangular lattice with anisotropic hopping on a cylinder of circumference 4 with YC boundary conditions (YC4 cylinder); dashed edges in the lattice indicate hopping strength $t'$ while solid edges indicate hopping strength $t$.  The gray lines at top and bottom are identified together to form a line running along the length of the cylinder.  These two distinct ways of orienting the anisotropic bonds on the cylinder we refer to as (a) ``symmetric'' and (b) ``asymmetric''.  {\bf (c)}-{\bf (e)} $U=0$ Fermi surface at half filling with $t'/t$ = 0.5, 1, and 2, respectively.  Allowed momentum cuts through the Brillouin zone for the YC4 cylinder are shown by solid horizontal lines for the case of symmetric anisotropy and dashed diagonal lines for asymmetric anisotropy.  (In the latter case, the Brillouin zone is actually rotated such that the dashed lines are horizontal.)\label{fig:lattices}}
\end{figure}

With the importance of anisotropy in mind, many theoretical studies have considered the Hubbard model or its strong-coupling limit, the Heisenberg model~\cite{MacDonald1988}, on the anisotropic triangular lattice. One of the three bonds in the lattice is chosen to be different from the other two; see Fig.~\ref{fig:lattices}.  These models thus interpolate from the square lattice in one limiting case to a set of uncoupled chains in the other with the isotropic lattice as a special point in between. 
The many theoretical works, using a diverse range of techniques such as variational Monte Carlo (VMC) simulations~\cite{Yunoki2006,Watanabe2006,Heidarian2009,Tocchio2009,Tocchio2013,Tocchio2014,Ghorbani2016}, exact diagonalization (ED)~\cite{Weng2006,Koretsune2007,Clay2008,Thesberg2014}, dynamical mean field theory (DMFT)~\cite{Kyung2006,Goto2016,Acheche2016}, variational cluster approximation (VCA)~\cite{Yamada2014,Yamada2014a,Laubach2015} and (for the Heisenberg model only) the density matrix renormalization group (DMRG)~\cite{Weng2006,Weichselbaum2011} and series expansions~\cite{Weihong1999,Pardini2008}, find a large variety of phases.  There is general agreement on some features, such as square lattice N\'{e}el ordering for a large portion of the phase diagram when one bond is weaker than the other two, and $120^\circ$ three-sublattice magnetic ordering in the strong coupling limit for the isotropic lattice. Furthermore, a wide range of studies agree on the presence of a nonmagnetic insulating (NMI) phase at intermediate interaction around isotropic hopping~\cite{Morita2002,Motrunich2005,Kyung2006,Sahebsara2008,Clay2008,Tocchio2009,Yoshioka2009,
Yang2010,Antipov2011,Tocchio2013,Laubach2015,Mishmash2015,Misumi2017,Shirakawa2017}. 
In the case where the distinct bond is stronger than the other two, there is much more disagreement, with proposed phases including collinear magnetic order, spiral magnetic order, and various spin liquids.
We provide a reasonably comprehensive review of these past works below.

In this paper, we provide the first DMRG study of the full Hubbard model on the anisotropic triangular lattice.  A primary motivation for this work is to further the understanding of the aforementioned NMI phase.  In a recent work, we identified the NMI, whose nature had not been determined numerically before, as a topologically ordered chiral spin liquid (CSL)~\cite{Kalmeyer1987,Szasz2020}; however, our previous study considered only the isotropic line, so a natural follow-up question is whether the CSL remains stable upon the introduction of anisotropy in the hopping. If the CSL does prove stable, that would suggest it should be taken seriously as a possible explanation of experimentally observed behavior; this is especially important to check because the possibility of a CSL was not investigated in past theoretical studies of the anisotropic model.  On the other hand, the CSL might give way to various other spin liquids with a small amount of anisotropy, in which case this study could reveal other candidate states to look for in experiments and to compare with other theoretical works.  

In addition to providing an independent perspective that can be compared with results from other numerical approaches, DMRG~\cite{White1992,White1993}, a variational algorithm for finding ground states within the matrix product state (MPS) ansatz~\cite{Ostlund1995,Schollwock2011}, has some crucial advantages.  In particular, 
DMRG calculations capture the full many-body correlation effects in the system, and the ansatz is not explicitly biased towards certain types of states such as spin liquids or magnetic orders.  
On the other hand, one key limitation of DMRG is that it is efficient in one dimension, but not in two dimensions.  Consequently, to study a two-dimensional model such as the Hubbard model on a triangular lattice, one must restrict the system to a quasi-one-dimensional system such as a cylinder of finite circumference; the calculation effort scales exponentially in the circumference, limiting simulations for spinful fermions to circumferences on the order of six lattice sites or fewer.  In order to make meaningful statements about the original two-dimensional model, we study four different cylinder geometries and compare the results; phases that consistently appear can be assumed to be present also in the two-dimensional limit, while phases appearing only for some cylinders should rather be viewed as possibilities which may or may not appear in the full two-dimensional model.

The phase diagram we find is broadly similar to the results of past works mentioned above.  In addition to the three phases found in our previous work on the isotropic model~\cite{Szasz2020}, namely the spiral magnetic order at large $U/t$, apparently metallic phase (which is likely a Luther-Emery liquid~\cite{Gannot2020}) at low $U/t$, and CSL in between, like the various past works we find that a large portion of the phase diagram for one weak bond is filled by the square lattice N\'{e}el order, and in the other limit of one strong bond we find a large variety of phases including the previously predicted collinear magnetic order and spin liquids, as well as some phases that have not been predicted before, such as phases with alternating orbital charge currents.  For the specific question of whether the CSL is stable to hopping anisotropy, we find that it remains the ground state with up to about 5 to 10\% anisotropy, beyond which we observe magnetic ordering as well as possible gapless spin liquids.

The remainder of this paper is organized as follows: In Sec.~\ref{sec:model} we introduce the model we study, including the specifics of the cylinder geometries we use.  We review the results of past theoretical work on both the Hubbard and Heisenberg models on the anisotropic triangular lattice in Sec.~\ref{sec:theory}, before summarizing our results in Sec.~\ref{sec:PD}; for each of the four cylinder geometries we study, we present phase diagrams in two parameters: coupling strength and degree of anisotropy.  In Sec.~\ref{sec:data}, we show the key data from our simulations that inform the phase diagrams; for interested readers, further data are included in the Supplemental Material~\cite{SuppMat}.  Finally, we conclude in Sec.~\ref{sec:discuss} with a discussion of how our results fit with both the past theoretical works and experimental findings.

\section{The Model\label{sec:model}}
We consider the Hubbard Hamiltonian,
\begin{equation}
H = -\sum_{\langle i j\rangle\s} t^{ }_{ij} c_{i\s}^\dg c^{ }_{j\s} + \text{H.c.} + U\sum_i n^{ }_{i\su}n^{ }_{i\sdn},\label{eq:Hubbard_H}
\end{equation}
where $c^{ }_{i\s}$ ($c_{i\s}^\dg$) is the fermion annihilation (creation) operator for spin $\s$ on site $i$  and $n=c^\dg c$ is the number operator.  $\langle\cdot\rangle$ indicates nearest neighbor pairs on the triangular lattice; on this lattice there are three distinct bonds, and we consider anisotropic hopping $t_{ij}^{ }$ such that two bonds have hopping strength $t$ and one has hopping strength $t'$, as shown in Figs.~\ref{fig:lattices}(a) and (b).  We work at half filling with net zero spin, so that $\sum_i\langle n^{ }_{i\su}\rangle = \sum_i\langle n^{ }_{i\sdn}\rangle = N/2$, where $N$ is the total number of sites.  In Figs.~\ref{fig:lattices}(c)-(e) we show the $U=0$ Fermi surface at half filling with $t'/t = 0.5$, 1, and $2$; the transition from a closed (two-dimensional) to open (quasi-one-dimensional) Fermi surface is at $t'/t \approx 1.636$.

To study this model using the DMRG method, we wrap the two-dimensional triangular lattice onto an infinitely long cylinder of finite circumference.  We use the so-called YC boundary conditions~\cite{Yan2011,Szasz2020}, for which the triangles are oriented such that one of the sides runs along the circumference of the cylinder.  Given the YC boundary conditions, there are two distinct ways of introducing the aforementioned anisotropy.  If the bonds with strength $t'$ run around the cylinder circumference, this preserves all spatial symmetries of the cylinder; we therefore refer to this orientation, shown in Fig.~\ref{fig:lattices}(a), as ``symmetric anisotropy.''  Conversely, if the $t'$ bonds are on one of the diagonal directions, as in Fig.~\ref{fig:lattices}(b), the mirror symmetries of the cylinder are broken; we refer to this orientation as ``asymmetric anisotropy.''

As in our paper on the isotropic model~\cite{Szasz2020}, we explicitly conserve the momentum quantum numbers associated with translation around the cylinder circumference by rewriting the Hamiltonian in a mixed real- and momentum-space basis with single-particle operators $c_{x, k_y, \sigma}$~\cite{Motruk2016, Ehlers2017}.  This improves the computational efficiency of our simulations and also allows us to separately find the ground state in different momentum sectors.

We focus particularly on the YC4 cylinder, with four sites around the circumference, which is simultaneously small enough to allow for relatively converged simulations and large enough to capture at least some behavior of the full two-dimensional model.  In order to better assess how representative the YC4 cylinder is of the 2D model, we also make a more limited study of both the YC3 and YC6 cylinders, revealing that some phases appear in all cases and are likely robust to the 2D limit while others are limited to just one of the cylinders and should be viewed only as candidates for existence in two dimensions.


\section{Past theoretical results\label{sec:theory}}

Before presenting the results of our simulations, we review past theoretical works on both the Hubbard and Heisenberg models on the anisotropic triangular lattice.  

In the case of the Heisenberg model, two observations are firmly established with a variety of methods. First, when one bond is weaker than the other two, the N\'eel order remains stable in a wide region extending from the square lattice limit, and second, the isotropic case exhibits $120^{\circ}$ spiral order~\cite{Huse1988,Capriotti1999,White2007}. Density matrix renormalization group (DMRG) calculations suggest a continuous variation of the angle of the spiral order from the isotropic point to both the N\'eel state and the uncoupled chain limit~\cite{Weichselbaum2011}. Variational Monte Carlo (VMC) simulations~\cite{Yunoki2006,Heidarian2009,Ghorbani2016}, resonating valence bond mean-field theory~\cite{Hayashi2007} and Schwinger boson theory~\cite{Gonzalez2020} on the other hand report the presence of at least one spin liquid phase towards the latter, a scenario that is partly supported by the detection of a magnetically disordered phase with collinear spin correlations in a functional renormalization group investigation~\cite{Reuther2011} and by earlier DMRG and ED studies~\cite{Weng2006}.
Furthermore, a VMC study puts forward the possibility of another spin liquid phase with competitive energy between the N\'eel and spiral orders, but the results remain inconclusive~\cite{Ghorbani2016}; this possibility is also seen in the Schwinger boson theory work~\cite{Gonzalez2020}.  Around the same region, early series expansion calculations have instead suggested a dimer ordered phase~\cite{Weihong1999}.
While the above numerical studies favor either incommensurate spiral order or a spin liquid in the weakly coupled chain case, collinear order has been proposed from renormalization group (RG)~\cite{Starykh2007,Ghamari2011} studies and shown to be strongly competitive using the coupled cluster method~\cite{Bishop2009} and ED~\cite{Thesberg2014}. In a series expansion, the spiral is shown to prevail over collinear order, but small magnetization and unclear convergence properties might hint at a spin liquid state in agreement with many of the numerical works~\cite{Pardini2008}.

Turning to the Hubbard model, in addition to the anisotropy, the ratio of interaction to kinetic energy $U/t$ represents another degree of freedom and adds to the complexity of the phase diagram; the range of open questions grows concomitantly larger.  In the isotropic case, numerous methods have established the existence of a nonmagnetic insulating (NMI) phase for intermediate interaction strength, however, the determination of its precise nature escaped these approaches~\cite{Morita2002,Kyung2006,Sahebsara2008,Clay2008,Yoshioka2009,Antipov2011,Laubach2015,Misumi2017,Shirakawa2017}. Based on VMC calculations on a Heisenberg model with ring exchange resulting from a $t/U$ expansion of the Hubbard model, a spin liquid with spinon Fermi surface (SFS) was long believed to be a strong candidate for the state~\cite{Motrunich2005}; this scenario also has some support from ED and DMRG simulations~\cite{Yang2010,Block2011,He2018}. Despite this model capturing the main effect of charge fluctuations near the Mott transition, it remained unclear whether the SFS would also appear in the Hubbard model. (In fact, the emergence of the state is still under debate even in the spin model~\cite{Aghaei2020}.) In our previous DMRG study, we determined the nature of the NMI phase in the full Hubbard model for the first time and suggested that it is a topologically ordered chiral spin liquid~\cite{Szasz2020,Kalmeyer1987}.  More recent papers, using a different variant of DMRG, also find the CSL~\cite{Zhu2020,Chen2021} and its presence was subsequently detected in the effective spin model as well~\cite{Cookmeyer2021}.

With the introduction of anisotropy, two limits are still very well understood: when $U=0$, the model is exactly solvable, and the system is metallic; when $t'=0$, the model reduces to the square lattice Hubbard model and has long-range N\'{e}el antiferromagnetic order for all $U>0$.  
As in the Heisenberg model, the square lattice N\'{e}el state extends through much of the $t'/t<1$ portion of the phase diagram, as reported in studies by numerous techniques~\cite{Morita2002,Watanabe2006,Kyung2006,Sahebsara2006,Koretsune2007,Watanabe2008,Tocchio2013,Yamada2014,Laubach2015,Goto2016}. VMC simulations suggest the presence of a spin liquid phase between the N\'eel order and the spiral phase around the isotropic line for large $U/t$, similar to the possibility in the Heisenberg model~\cite{Tocchio2009,Tocchio2013}. Moreover, a $d$-wave superconducting state has been proposed to appear at lower $U/t$ between the metallic and N\'eel phases from cellular DMFT (CDMFT)~\cite{Kyung2006}, variational cluster perturbation theory~\cite{Sahebsara2006}, and some VMC~\cite{Watanabe2006} calculations.  Conversely, we note the absence of a superconducting state in other VMC studies~\cite{Watanabe2008,Tocchio2013}, and in path integral renormalization group~\cite{Morita2002}, ED~\cite{Koretsune2007}, variational cluster approximation (VCA)~\cite{Yamada2014,Laubach2015}, and complementary DMFT~\cite{Goto2016} studies. 

The low-$U$ phase near the isotropic line is mostly referred to as (paramagnetic) metal in the studies we review here, however, this might not be entirely correct. The noninteracting state at $U/t=0$ is obviously a metal, but RG studies in the full two-dimensional model have shown that it becomes a $d+id$ superconductor for infinitesimally small interaction~\cite{Raghu2010,Nandkishore2014}. If there is no other phase transition between the low-$U$ state and the NMI state, this full region of the phase diagram could be a superconductor, but possibly with an exponentially small gap which would make it likely to appear to be a metal in many numerical studies.  There is also evidence for similar behavior at low $U/t$ on the cylinder geometries relevant for DMRG simulations~\cite{Gannot2020}; the phase would then be a Luther-Emery liquid (LEL)~\cite{Luther1974}. On the numerical side, a modified version of the VCA from Ref.~\cite{Laubach2015} suggests that a $d+id$ superconducting state is lower in energy than the metal. This superconductor scenario would also lead to a more natural description of the transition to the CSL.

The situation in the case of hopping anisotropy towards the 1D chain limit with $t'/t > 1$ has been much less studied. Both VMC~\cite{Tocchio2014} and VCA~\cite{Yamada2014,Yamada2014a} calculations report the presence of a collinear magnetically ordered phase at intermediate $U/t$ and the spin liquid at large $U/t$ connected to the one suggested in the Heisenberg model. The occurence of the collinear state is in agreement with ED and CDMFT studies, which, however, do not find indications of the spin liquid~\cite{Clay2008,Acheche2016}.


\section{Phase diagram summary\label{sec:PD}}

In this section, we present a summary of the phase diagram we observe for each of four different setups: YC4 cylinder with symmetric anisotropy, YC4 cylinder with asymmetric anisotropy, YC3 cylinder with symmetric anisotropy, and YC6 cylinder with symmetric anisotropy.  Each summary phase diagram is shown in  Fig.~\ref{fig:summary_PDs}, with the supporting data presented below, in Sec.~\ref{sec:data}.

\begin{figure*}
\includegraphics[width = \textwidth]{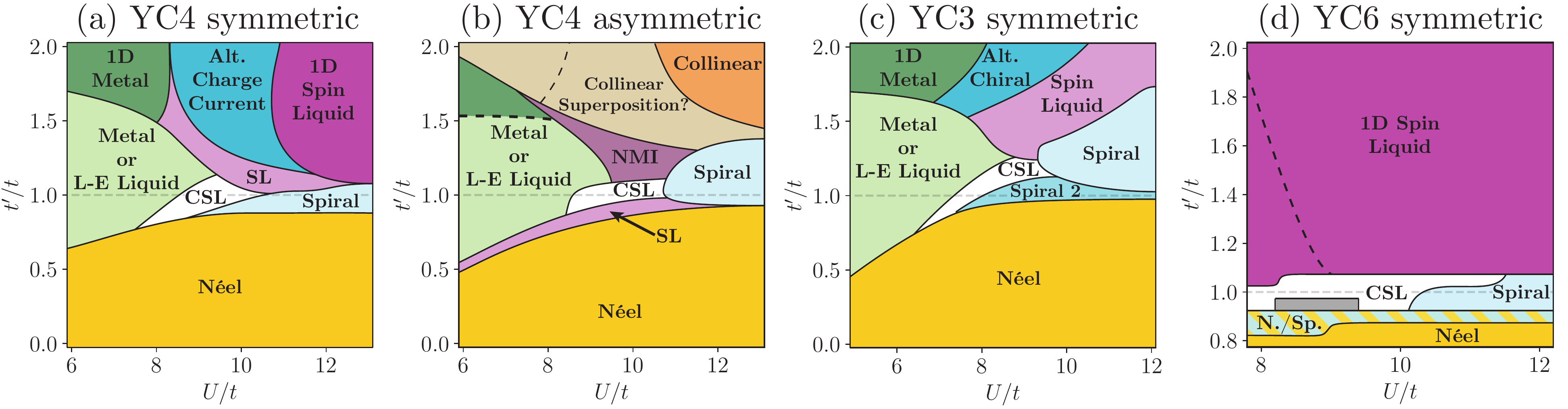}
\caption{Summary phase diagrams for four different cylinder geometries.  {\bf (a)} YC4, symmetric anisotropy; {\bf (b)} YC4, asymmetric anisotropy; {\bf (c)} YC3, symmetric anisotropy; {\bf (d)} YC6, symmetric anisotropy.  Note that the parameter ranges are not all the same.  We briefly describe the observed phases (in alphabetical order by label): 
\emph{\bf 1D Metal}---phase with a large charge 1 correlation length like a metal and with an open (quasi-one-dimensional) Fermi surface;
\emph{\bf 1D Spin Liquid}---a nonmagnetic insulator with some one-dimensional character (e.g. in the spin structure factor), and which has isolated gapless points;
\emph{\bf Alt.~Charge Current}---a phase with charge currents running around the cylinder circumference, in opposite directions on neighboring rings, and which also has a nonzero scalar chiral order parameter with alternating sign; 
\emph{\bf Alt.~Chiral}---a similar phase, but with currents and chiral order parameters varying with a 4 ring unit cell; 
\emph{\bf Collinear}---a state with antiferromagnetic ordering along the distinct (strong) bond, and with ferromagnetic and antiferromagnetic ordering respectively along the other two bonds;
\emph{\bf Collinear superposition}---possibly a superposition of collinear order in two different orientations;
\emph{\bf CSL}---chiral spin liquid; 
\emph{\bf Metal or L-E Liquid}---phase that appears metallic in our data, but a large charge 2 correlation length and RG theory~\cite{Gannot2020} suggest is a Luther-Emery liquid; 
\emph{\bf N\'{e}el}---square lattice antiferromagnetic spin order; 
\emph{\bf NMI}---nonmagnetic insulator, which appears to be gapped;
\emph{\bf SL} or \emph{\bf Spin Liquid}---a nonmagnetic insulating phase that appears to be spin gapless;
\emph{\bf Spiral}---spiral magnetic order; 
\emph{\bf Spiral 2}---part of the same spiral-ordered phase in two dimensions, but distinct on the cylinder.
Dashed lines denote locations where there may be a phase transition, but our data are not conclusive.
Finally, we note two regions for YC6 in which the correct phase is not clear at accessible MPS bond dimensions: the striped region for YC6, labeled N./Sp. (N\'{e}el/Spiral), which may belong to either phase, and the small gray region, which may belong to the CSL or spiral order.  In both cases, different momentum sectors show different behavior, and the energy difference is comparable to or smaller than the MPS truncation error. 
\label{fig:summary_PDs}}
\end{figure*}

Of course, the real aim is to determine the phase diagram of the model on the full two-dimensional lattice, not the phases that appear on certain finite-circumference cylinders.  While a rigorous conclusion about the two-dimensional model is not possible from our data, we can make some progress, noting that:
\begin{enumerate}
\item Any phase where all four phase diagrams agree is more likely to be present in the two-dimensional model.
\item Larger cylinders are more representative of the two-dimensional model in principle, but they also require a much larger MPS bond dimension to converge; when the bond dimension is too small, the wrong state may be energetically favored.  

Some intuition for the question of how small of a circumference is ``too small'' can be gained by comparing the YC3 phase diagram with those of the larger cylinders.  Evidently, the whole phase diagram is shifted upwards, so that $t'/t\approx 1.2$ acts like the isotropic line, meaning that correlations around the cylinder circumference introduce an effective anisotropy.  So at least near $t'/t=1$, YC4 may be large enough to capture two-dimensional behavior, while YC3 might not be.

\item Because the symmetric anisotropy case has the same mirror symmetries as the full two-dimensional anisotropic lattice, it seems likely to better represent the true two-dimensional ground state.  Some evidence for this can be found in Ref.~\cite{Acheche2016}.
\item On the other hand, in the limit of large $t'/t$, the case of symmetric anisotropy may act as a single weakly coupled one-dimensional chain, with each effective site being one ring of the cylinder; in contrast, asymmetric anisotropy gives a small number of weakly coupled infinite one-dimensional chains.  The latter may be a more natural way to study the two-dimensional large-$t'$ limit of weakly coupled one-dimensional chains.
\item Specific types of magnetic ordering are stabilized/destabilized by whether they are commensurate with the cylinder circumference. 
\end{enumerate}

\noindent With these points in mind, we can make some predictions about what the phase diagram of the full two-dimensional model should look like.  The metal/LEL phase, CSL, and spiral order should appear on the isotropic line, as in our previous work~\cite{Szasz2020}.  For $t'/t < 1$, the majority of the phase diagram is taken up by the N\'{e}el order, and there may be an additional phase between the N\'{e}el phase and the CSL and spiral phases near $t'/t=1$.  However, this additional phase is clearly observed only for YC4 with asymmetric anisotropy, which breaks extra symmetries.  There is an extra phase in this region for the YC3 cylinder as well, but that appears to also correspond to the spiral phase of the two-dimensional model, appearing as distinct only due to the finite circumference.  Thus overall our data weakly suggest the absence of such an additional phase in two dimensions.  

For $t'/t > 1$, any predictions are much less certain.  At low $U/t$, we consistently find what we call the ``1D metal'' phase, for which the longest correlation lengths are for charge 1 excitations, and where the $k$-space occupation numbers are consistent with an open Fermi surface; for YC4 and YC3 with symmetric anisotropy, this phase is clearly distinct from the metal/LEL, but for YC4 with asymmetric anisotropy, it is distinguishable only via the occupation numbers, so its status as a truly distinct phase is not entirely clear.  

As $U$ increases, we find multiple spin liquids, collinear magnetic order, and time-reversal symmetry-breaking phases whose scalar chiral order parameter varies in sign and/or magnitude between triangular plaquettes.  The latter phases appear only for YC3, the smallest and least two-dimensional of the cylinders, and YC4 with symmetric anisotropy, in the $t'/t\gg 1$ limit where the asymmetric anisotropy may be more representative.  Consequently, these phases are unlikely to be present in the two-dimensional model.  We also note that the spiral magnetic order may extend to larger $t'/t$ in two dimensions, since general spiral orders are incommensurate on finite-circumference cylinders and thus might be artificially disfavored.  This incommensurability can be partially addressed by flux insertion, as we discuss in the Supplemental Material, Sec.~II.A.5., but our data from such calculations are not conclusive.

The spin liquids we find seem to be of two types.  First is the ``1D spin liquid'' predicted in references \cite{Yunoki2006,Heidarian2009,Tocchio2014,Ghorbani2016,Gonzalez2020}, which we observe in the YC4 symmetric case and also for the YC6 cylinder.  Despite the name, this is not truly a one-dimensional phase.  While some properties, such as spin-spin correlations, indeed appear very one-dimensional, the excitation spectrum as revealed by MPS transfer matrix spectra with flux insertion has only isolated gapless points in the two-dimensional Brillouin zone consistent with the characterization from VMC.  The second spin liquid is closer to the isotropic line and is found by gapping out charge from the 1D metal phase.  Both types of spin liquids, as well as the collinear magnetic order, are strong candidates for the $t'/t>1$ regime for the full two-dimensional model.


\section{Data for different cylinders\label{sec:data}}

In this section, we present the most important pieces of data that lead to the phase diagram summaries of Fig.~\ref{fig:summary_PDs}.  In particular, we show the following observables:
\begin{itemize}[label={},leftmargin=*,listparindent=1em]
\item \emph{Magnetic ordering}, as measured by the spin structure factor.  For each MPS wavefunction, we compute $\langle S_z S_z\rangle$ correlations to a distance of six sites along the cylinder, then Fourier transform to find an approximation to the spin structure factor.  We confirm using a random subset of parameter points both that (a) spin rotation symmetry is preserved in the simulation, so that $\langle S_z S_z\rangle = \langle S_x S_x\rangle = \langle S_y S_y\rangle$, and (b) the range of six sites is sufficient to capture the full spin structure factor, either because there is true long-range order which is already clear at this distance or because long-range order is prevented since it would require spontaneously breaking a continuous symmetry in one dimension. In the latter case, the decay is fast enough that farther correlations do not contribute significantly to the structure factor.

We illustrate the measured spin structure factors in two ways: by showing the structure factor at a representative point in each phase, and by showing its height at specific points in the Brillouin zone (e.g. $M$ and $K$ points) for all points in the phase diagram.

These spin structure factor results can be compared with the expectations for several types of magnetic ordering, including square lattice N\'{e}el order, spiral order, and collinear order; for details on how these different orderings are expected to manifest on the cylinders we study, see the Supplemental Material~\cite{SuppMat}, Sec.~I.

\item \emph{Time-reversal symmetry breaking}, as measured by the scalar chiral order parameter, $\langle \mathbf{S}_i\cdot(\mathbf{S}_j\times\mathbf{S}_k)\rangle$ where $i$, $j$, $k$ label the three vertices of a triangular plaquette in the lattice.  This is nonzero in the chiral spin liquid phase that appears near the isotropic line, and we also observe other phases with nonzero chiral order on some cylinders in the large $t'/t$ limit.

\item \emph{Correlation lengths} for excitations with charge 0, 1, and 2.  These serve as a rough indication of whether phases are gapped or gapless in these charge sectors.  Ideally, one would instead perform a finite entanglement scaling calculation~\cite{Tagliacozzo2008,Pollmann2009,Pirvu2012}, but for a two-dimensional phase diagram this requires more computation time than is feasible, and, as we show below, the correlation lengths already provide clear intuition.
\end{itemize}
While these data are sufficient to distinguish different phases, and in some cases to clearly identify them, we also consider a variety of other quantities that help to identify the phases, including entanglement entropy and entanglement spectra~\cite{Li2008,Qi2012}, transfer matrix spectra~\cite{Zauner2015,Szasz2020}, and occupation in the Brillouin zone, among others.  All of these additional data can be found in the Supplemental Material~\cite{SuppMat}.


\subsection{YC4, symmetric anisotropy}
\begin{figure*}
\includegraphics[width = \textwidth]{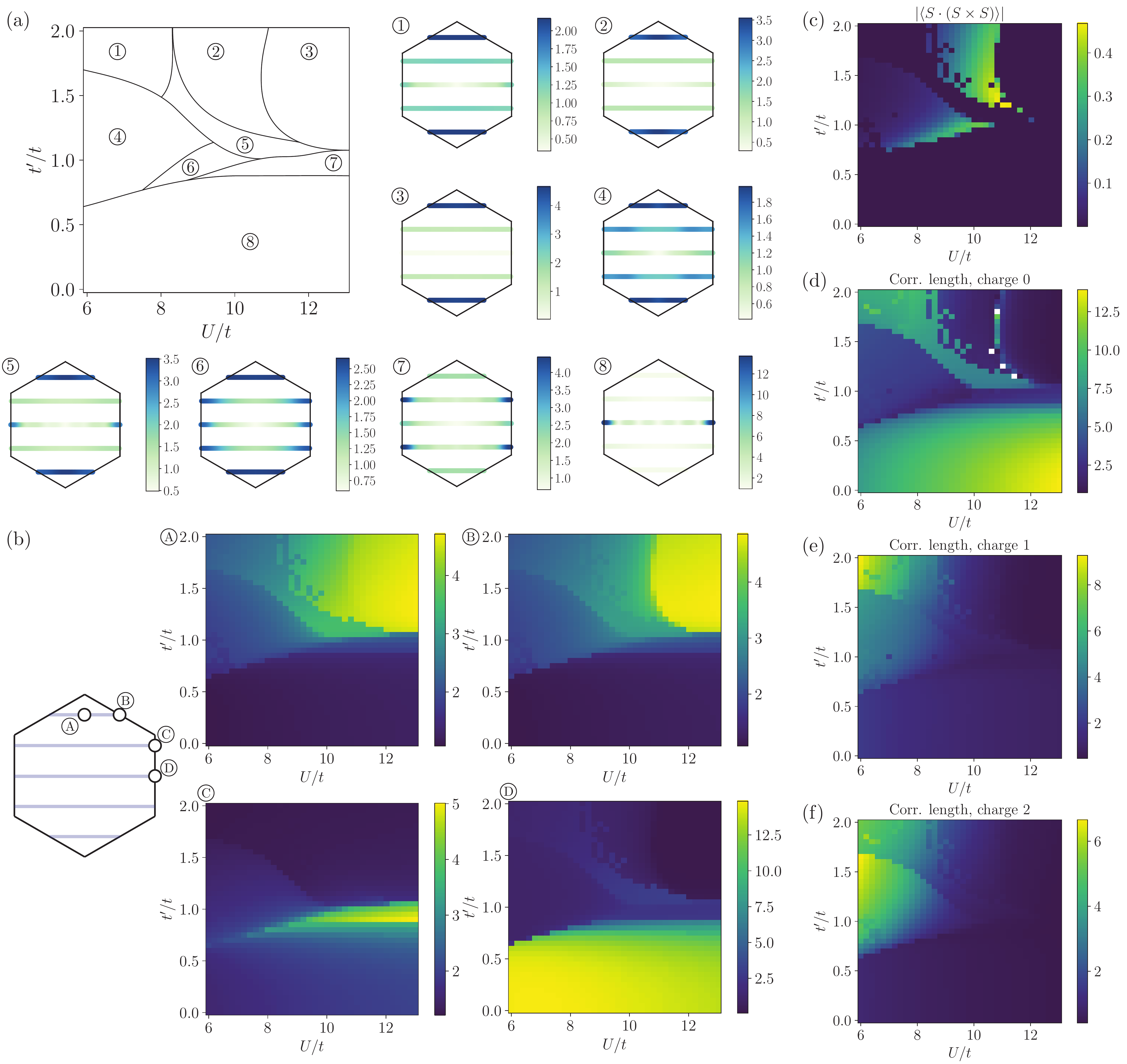}
\caption{Data for YC4 cylinder with symmetric anisotropy, with MPS bond dimension $\chi=4000$.  {\bf (a)} Spin structure factor, view 1.  We show the spin structure factor computed on allowed momentum cuts at one representative point in each of the eight phases we observe.  {\bf (b)} Spin structure factor, view 2.  We show, for the full range of $t'/t$ and $U/t$, the value of the spin structure factor at the four points labeled \Circled{A}, \Circled{B}, \Circled{C}, and \Circled{D}.  The $M$ points \Circled{B} and \Circled{D} correspond to collinear and N\'{e}el order respectively.  The point \Circled{C} is as close as possible to the $K$ points, where the $120^\circ$ order appears, on the allowed momentum cuts.  The point \Circled{A} is included because a comparison with \Circled{B} distinguishes between collinear order and a ``one-dimensional'' order where the whole line containing those points is a sub-extensive peak. {\bf (c)} Scalar chiral order parameter $\langle \mathbf{S}\cdot(\mathbf{S}\times\mathbf{S})\rangle$.  Note that the nonzero value at low $U/t$ is a finite bond dimension effect; see Fig.~3(e) of Ref.~\cite{Szasz2020}.  In the CSL, around the isotropic line, the order parameter has the same sign on every triangular plaquette.  In the large $t'/t$ phase where the order parameter is nonzero, the sign alternates on neighboring rings; see the Supplemental Material~\cite{SuppMat} for an illustration of the pattern.   {\bf (d)} Correlation length for charge 0 excitations (i.e. spin excitations), computed using the MPS transfer matrix.  A large correlation length will be found if spin excitations are gapless.  {\bf (e)} Correlation length for charge 1 excitations.  A large correlation length is implied by gapless charge excitations. {\bf (f)} Correlation length for charge 2 excitations.  A large correlation length may indicate a tendency towards superconductivity.
\label{fig:YC4_sym_data}}
\end{figure*}

Data for the YC4 cylinder with symmetric anisotropy are shown in Fig.~\ref{fig:YC4_sym_data}; to be precise, we show the spin structure factor, scalar chiral order parameter, and correlation length in each of several charge sectors.  These quantities are calculated from MPS ground states with bond dimension $\chi=4000$, using unit cells of one, two, and three rings, and initialized with different conserved momenta, including total momentum around the cylinder of 0 for each unit cell size as well as momentum $\pi$ for one and two ring unit cells.  The three ring unit cell data were computed by adiabatically increasing or decreasing anisotropy starting from the isotropic point, while the others were computed directly at each point in the phase diagram; both approaches yield the same observed phases.  At each point in the phase diagram, the figures show measured quantities using the state with the lowest energy among the various datasets.  We have also performed the same simulations with bond dimensions of 1000 and 2000, finding the same phases but with somewhat shifted boundaries, especially at low $U$.  Some representative figures for these lower bond dimensions are available in the Supplemental Material, along with various additional quantities at bond dimension 4000: real-space spin correlations, $k$-space occupation numbers, 
transfer matrix and entanglement spectra, and various properties as a function of spin flux insertion in the high-$U$ limit~\cite{SuppMat}.

Each phase can be at least partially identified using these data.  We now discuss the evidence for each phase, roughly in order from most to least clearly identified from our data.

\begin{itemize}[label={},leftmargin=*,listparindent=1em]
\item \emph{N\'{e}el order}: The most obvious phase, and also the one occupying the largest portion of the phase diagram, is the square lattice N\'{e}el magnetic order, labeled as \Circled{8} in Fig.~\ref{fig:YC4_sym_data}(a).  Referring to Fig.~\ref{fig:lattices}(a), in the limit $t'/t=0$ the dashed line vanishes entirely and the lattice becomes unfrustrated, leading to magnetic ordering that is ferromagnetic within each ring and antiferromagnetic between rings.  In the spin structure factor, this is indicated by extensive peaks at the $k_y=0$ $M$ points, exactly as we see in Fig.~\ref{fig:YC4_sym_data}(a)\Circled{8}.  The energy is lowest with a two ring unit cell, in which case the translation symmetry of the Hamiltonian is spontaneously broken in the simulation: there is a net spin up on half the rings and net spin down on the other half.\footnote{Note that this phase consequently breaks spin rotation symmetry; since a continuous symmetry cannot be spontaneously broken in one dimension, this must be a finite bond dimension effect.  Indeed, the degree of symmetry breaking is reduced with increasing MPS bond dimension, though a reliable extrapolation is not possible from our data.}
With a one or three ring unit cell, so that the translation symmetry breaking is not allowed, the N\'{e}el order still appears in the correlation functions, but the correlation strength decays slowly to zero along the cylinder.

\item \emph{Spiral order}: The spiral-ordered phase, labeled by \Circled{7} in Fig.~\ref{fig:YC4_sym_data}(a) and exemplified by the $120^\circ$ three-sublattice order at the isotropic point, is identifiable primarily via the spin structure factor.  For the special case of $120^\circ$ order, on the full two-dimensional lattice the structure factor would have peaks at the $K$ points, the corners of the Brillouin zone; we look for peaks at the closest allowed momenta, including point \Circled{C} in Fig.~\ref{fig:YC4_sym_data}(b).  Based on analysis of the classical Heisenberg model~\cite{Merino1999} as discussed in the Supplemental Material~\cite{SuppMat}, away from the isotropic point incommensurate spiral order is expected with peaks along the line connecting points \Circled{D} (for $t'/t = 1/\sqrt{2}$) and \Circled{C}, and continuing through the $K$ point to the bottom of the Brillouin zone, ending at the mirror of point \Circled{A} (for $t'/t\rightarrow\infty$).  However, in our data, we find that the spiral order is stabilized only in the vicinity of the isotropic line.  It would be reasonable to suppose that the spiral-ordered phase would shift with spin flux insertion, which can shift the allowed momentum cuts through the Brillouin zone and thus change which magnetic orders are effectively commensurate on the cylinder.  Indeed we see that spiral order near the isotropic line becomes stronger when the allowed momentum cuts include the $K$ points.  At small nonzero flux the spiral phase appears larger than at 0 flux, but phase boundaries are not clearly defined (see Supplemental Material~\cite{SuppMat}, Sec.~II.A.5.).

\item \emph{Chiral spin liquid}: The CSL, labeled by \Circled{6} in Fig.~\ref{fig:YC4_sym_data}(a), is most easily identifiable via the scalar chiral order parameter in Fig.~\ref{fig:YC4_sym_data}(c) and corresponds roughly to the green region near the isotropic line in that figure.  

\item \emph{Metal} or \emph{Luther-Emery liquid (LEL)}: This phase, labeled by \Circled{4} in Fig.~\ref{fig:YC4_sym_data}(a), was identified as a metal in our previous work~\cite{Szasz2020}, on the basis partly of finite entanglement scaling calculations giving a central charge matching the expectation for the $U=0$ state on the isotropic line.  More recent theoretical work using renormalization group calculations~\cite{Gannot2020} suggests that this phase is instead a LEL, which would not have been distinguishable from the metal at the bond dimensions we considered.  However, one feature of the data in our previous paper that is consistent with the LEL is a relatively large correlation length for charge 2 excitations, which may indicate a tendency towards superconductivity.  In Fig.~\ref{fig:YC4_sym_data}(f), we see that the charge 2 correlation length is indeed even larger in the parts of this phase with larger $t'/t$, which is a further confirmation that this phase may indeed be the LEL rather than a metal.  To conclusively demonstrate this fact, however, would require much larger bond dimensions than are currently accessible, in order to resolve very small gaps and correspondingly large correlation lengths.  If such data were available, finite entanglement scaling for the LEL should show a central charge that decreases from 6 to 1 with increasing bond dimension~\cite{Gannot2020}.

\item \emph{1D metal}: The upper left phase, labeled by \Circled{1} in Fig.~\ref{fig:YC4_sym_data}(a), can be seen from Figures \ref{fig:YC4_sym_data}(d) and (e) to have large correlation lengths for both charge and spin excitations, indicating that it is likely gapless.  The boundary with the phase below is between $t'/t=1.65$ and 1.7 (for bond dimension $\chi=4000$ as shown in the figure---it is at a slightly lower value of $t'/t$ for smaller bond dimensions~\cite{SuppMat}), which seems to correspond to the $U=0$ Fermi surface opening at $t'/t\approx 1.64$; indeed, we directly observe from $k$-space occupation numbers that this is the case, as shown in the Supplemental Material~\cite{SuppMat}, Fig.~S10. While this Fermi surface opening informs the label we assign to the phase, we emphasize that it is identifiable as a distinct phase from the LEL via other signatures, such as the aforementioned correlation lengths.   In particular, the charge 2 correlation length is much shorter in this phase, perhaps indicating that it is truly metallic, unlike the LEL.

\item \emph{Alternating charge current}: The upper middle phase, labeled by \Circled{2} in Fig.~\ref{fig:YC4_sym_data}(a), is identifiable in Fig.~\ref{fig:YC4_sym_data}(c) by a nonzero scalar chiral order parameter.  What is not clear from the figure, where we plot only the absolute value of the scalar chirality, is that in this phase the sign alternates on different plaquettes---the precise pattern is shown in Fig.~S11 of the Supplemental Material~\cite{SuppMat}.  Evidently, the phase breaks translation symmetry.  Accordingly, this phase is only observed when we use a two ring unit cell in the simulation; for one or three ring unit cells, the simulations do not converge well in this region of phase space.  The translation symmetry-breaking also manifests in the form of charge currents that run around the cylinder circumference, in opposite directions on neighboring rings, which is also discussed in the Supplemental Material~\cite{SuppMat}.  We emphasize that these are local orbital currents only, and there is no net current through the system. Note that the phase is charge gapped, as can be seen from the very small charge correlation length in Fig.~\ref{fig:YC4_sym_data}(e).

\item \emph{1D spin liquid}: The upper right phase, labeled by \Circled{3} in Fig.~\ref{fig:YC4_sym_data}(a), is most clearly identifiable by a peak in the spin structure factor that is uniform along $k_x$ at $k_y=\pi$, in other words by strong antiferromagnetic correlations around the cylinder and almost no correlations along the cylinder.  
However, the picture of the phase as one-dimensional breaks down when we perform flux insertion by twisting the periodic boundary conditions around the cylinder, which has the effect of shifting the allowed momentum cuts through the Brillouin zone; flux insertion data are provided in the Supplemental Material~\cite{SuppMat}.  We observe gapless points with flux insertion, at which the local magnetic ordering becomes much more correlated along the cylinder.  Our results for this phase are generally consistent with the phase called ``1D spin liquid'' in Refs.~\cite{Yunoki2006,Heidarian2009,Ghorbani2016,Tocchio2014,Gonzalez2020}. In particular, the structure factor agrees with the ones presented in Refs.~\cite{Heidarian2009,Tocchio2014} for this phase.  Although, as revealed by our flux insertion calculations, the phase is not truly one-dimensional, we also adopt the ``1D spin liquid'' name for consistency with the literature.

\item \emph{Gapless spin liquid}: Finally, the central phase labeled by \Circled{5} in Fig.~\ref{fig:YC4_sym_data}(a) appears to be a gapless spin liquid: there is no strong magnetic order, and comparing panels (d) and (e), there is evidently a transition from the 1D metal at which the charge correlation length becomes short, indicating a gap, while the spin correlation length is unchanged.  To distinguish between different gapless spin liquids, we again perform flux insertion, but the results are inconclusive.  Looking at the transfer matrix spectrum (Fig.~S12 of the Supplemental Material\cite{SuppMat}), we see that the gapless region of the Brillouin zone is small, possibly consistent with a nodal gapless spin liquid such as a Dirac spin liquid (DSL).  A more precise identification would probably require much larger bond dimensions and, due to shifting phase boundaries with flux insertion~\cite{Szasz2020}, flux insertion over a possibly large portion of the phase diagram, so we leave this question for future study.

\end{itemize}

\begin{figure*}
\includegraphics[width = \textwidth]{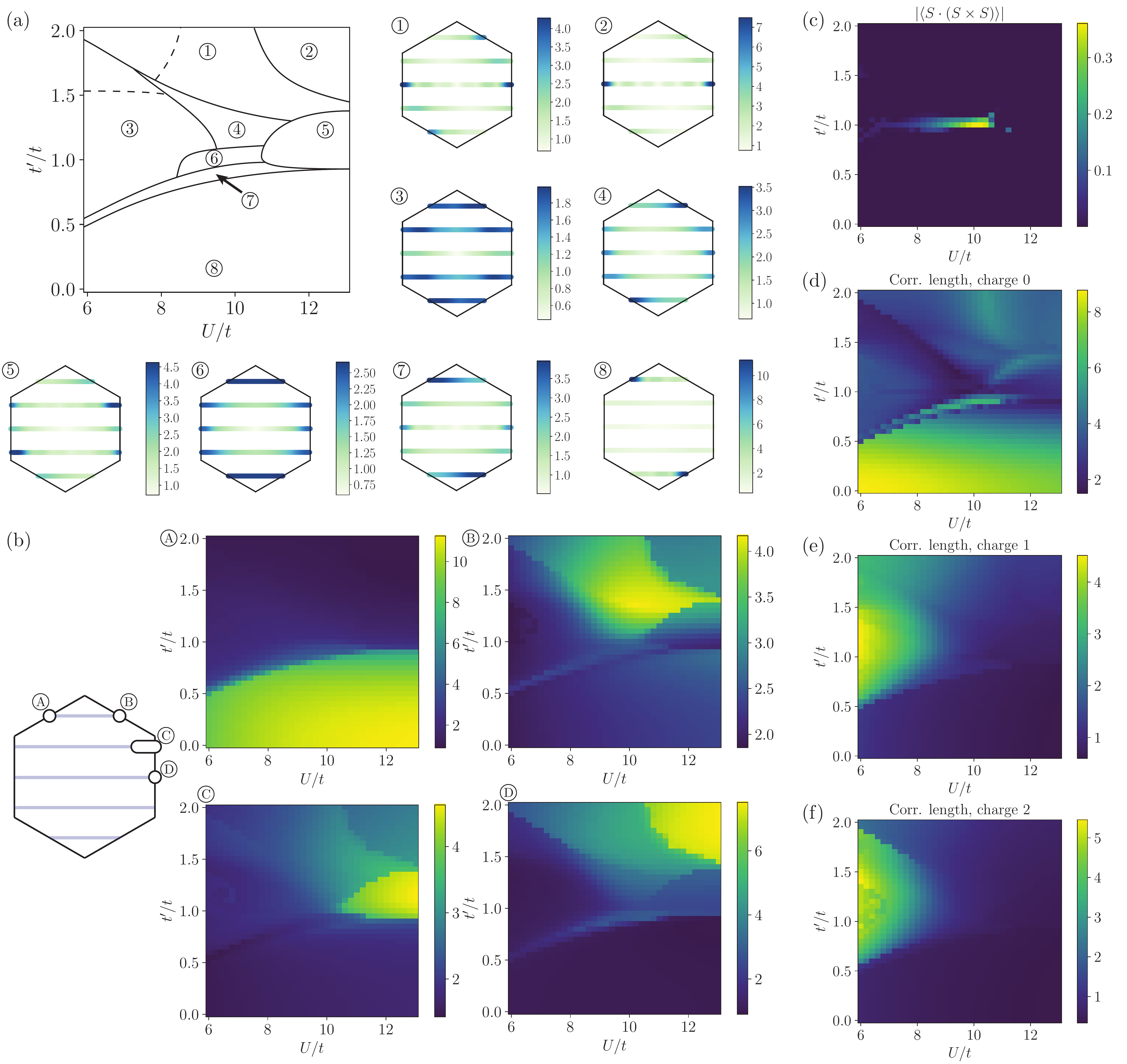}
\caption{Data for YC4 cylinder with asymmetric anisotropy, with MPS bond dimension $\chi=4000$.  {\bf (a)} Spin structure factor, view 1.  We show the spin structure factor computed on allowed momentum cuts at one representative point in each of the eight phases we observe. Note that for \Circled{3} and \Circled{5}, we have chosen points off the isotropic line for generality.     {\bf (b)} Spin structure factor, view 2.  We show, for the full range of $t'/t$ and $U/t$, the value of the spin structure factor at the three inequivalent $M$ points in the Brillouin zone, labeled \Circled{A}, \Circled{B}, and \Circled{D}, as well as the maximum height in the region labeled by \Circled{C}.  The $M$ point \Circled{A} corresponds to N\'{e}el order, while \Circled{B} and \Circled{D} correspond to possible collinear orders.  The region \Circled{C} includes all points on the $k_y=\pi/2$ momentum cut that are as close as possible to the expected peak locations for spiral order. {\bf (c)} Scalar chiral order parameter $\langle \mathbf{S}\cdot(\mathbf{S}\times\mathbf{S})\rangle$.  {\bf (d)} Correlation length for charge 0 excitations (i.e. spin excitations), computed using the MPS transfer matrix.  A large correlation length will be found if spin excitations are gapless.  {\bf (e)} Correlation length for charge 1 excitations.  A large correlation length is implied by gapless charge excitations. {\bf (f)} Correlation length for charge 2 excitations, possibly indicating a tendency towards superconductivity.
\label{fig:YC4_asym_data}}
\end{figure*}


\subsection{YC4, asymmetric anisotropy}

Data for the YC4 cylinder with asymmetric anisotropy are shown in Fig.~\ref{fig:YC4_asym_data}, again calculated from MPS ground states with bond dimension $\chi=4000$ and with one, two, and three ring unit cells, initialized with total momentum around the cylinder of $\pi$, 0 and $\pi$, and 0 per unit cell, respectively.  Lower bond dimension data are again available in the Supplemental Material, along with various other quantities.\cite{SuppMat}


Using these data, we identify the following phases:

\begin{itemize}[label={},leftmargin=*,listparindent=1em]

\item \emph{N\'{e}el order}: The bottom phase, labeled as \Circled{8} in Fig.~\ref{fig:YC4_asym_data}(a), is again square lattice N\'{e}el order, though the ordering pattern is rotated clockwise by $60^\circ$ compared to the symmetric case.  Here the bond that vanishes in the $t'/t=0$ limit is one of the diagonals, so the real-space spin correlations are ferromagnetic along that diagonal, leading to two differences from the symmetric case: the corresponding peak in the structure factor is at a different $M$ point, and translation symmetry is not broken.

\item \emph{Spiral order}: The spiral-ordered phase, labeled by \Circled{5} in Fig.~\ref{fig:YC4_asym_data}(a), is again identifiable from the spin structure factor.  It extends to larger $t'/t$ compared with the case of symmetric anisotropy, likely because in this case the expected peaks for spiral order in two dimensions are closer to the allowed momentum cuts at $k_y=\pm\pi/2$ when $t'/t>1$, while for the symmetric case they become farther from these cuts.

\item \emph{Chiral spin liquid}: The CSL, labeled by \Circled{6} in Fig.~\ref{fig:YC4_asym_data}(a), is again clearly identified by the scalar chiral order parameter, Fig.~\ref{fig:YC4_asym_data}(c).  With asymmetric anisotropy, this is the only phase with time-reversal symmetry breaking.

\item \emph{Metal} or \emph{Luther-Emery liquid}: This phase, labeled by \Circled{3} in Fig.~\ref{fig:YC4_asym_data}(a), is identified by a large correlation length for both charge 1 and charge 2 excitations, as shown in Figures \ref{fig:YC4_asym_data}(e) and (f).  The dashed line near the top shows where $k$-space occupation numbers seem to indicate an opening of the Fermi surface.  Unlike the case of symmetric anisotropy, no other quantities clearly change at this location, so we do not identify it as a separate phase.

\item \emph{Collinear order}: This phase, labeled by \Circled{2} in Fig.~\ref{fig:YC4_asym_data}(a), has collinear magnetic order, which in the present case of asymmetric anisotropy is nearly identical to that observed in the N\'{e}el phase with symmetric anisotropy, though weaker. Here too, the translation symmetry is spontaneously broken, with a net spin up on half the rings and a net spin down on the other half.  This ordering arises because the spins align antiferromagnetically along the strong (diagonal) bond.  In two dimensions, the other two bonds would be equivalent so that the collinear order, with ferromagnetic ordering along one of these directions and antiferromagnetic along the other, would have to spontaneously break a mirror symmetry or appear in a superposition of both orderings.  Here, the cylinder geometry already breaks the symmetry, and evidently the effective bond strength is larger on the remaining diagonal than around the cylinder, so the latter orders ferromagnetically.

\item \emph{Spin liquid}: This phase, labeled by \Circled{7} in Fig.~\ref{fig:YC4_asym_data}(a), appears to be a gapless spin liquid, much like the one for $t'/t > 1$ with symmetric anisotropy.  The likely gapless nature of the spin excitations is indicated by a large correlation length in Fig.~\ref{fig:YC4_asym_data}(d), while charge 1 and 2 excitations are clearly gapped, per Figures \ref{fig:YC4_asym_data}(e) and (f).  We perform spin flux insertion and use the transfer matrix spectrum to check the possible locations of spin-gapless points in the Brillouin zone, again finding a small gapless region that is possibly consistent with a DSL or other nodal gapless spin liquid.

\item \emph{Collinear superposition?}: This phase, labeled by \Circled{1} in Fig.~\ref{fig:YC4_asym_data}(a), is challenging to identify.  Like the collinear phase, it breaks translation symmetry, though the magnitudes of $\langle S_z\rangle$ on each ring are smaller.  Accordingly, the structure factor has peaks at the $k_y=0$ $M$ points.  However, there are also peaks at the $M$ points labeled by \Circled{B} in Fig.~\ref{fig:YC4_asym_data}(b), which correspond to the other possible collinear order after antiferromagnetic correlations along the strong bond have been fixed.  So one strong possibility is that this phase is a superposition of the two possible collinear orders.

We also note that there is a moderately large charge correlation length [see Fig.~\ref{fig:YC4_asym_data}(e)] in the leftmost part of the phase, which may indicate that there are actually two phases here, one which is metallic with an open Fermi surface, the other being the phase described in the previous paragraph.  As this is unclear, we denote this possible boundary by a dashed line.

\item \emph{Nonmagnetic insulator}: Finally, the phase labeled by \Circled{4} in Fig.~\ref{fig:YC4_asym_data}(a) appears to be some kind of nonmagnetic insulator, with no long correlation lengths and with minimal magnetic ordering even at short range.  It is sharply distinguished from the phase labeled \Circled{1} in \ref{fig:YC4_asym_data}(a) by the lack of translation symmetry breaking but seems otherwise similar.

\end{itemize}


\subsection{YC3, symmetric anisotropy}

\begin{figure*}
\includegraphics[width = \textwidth]{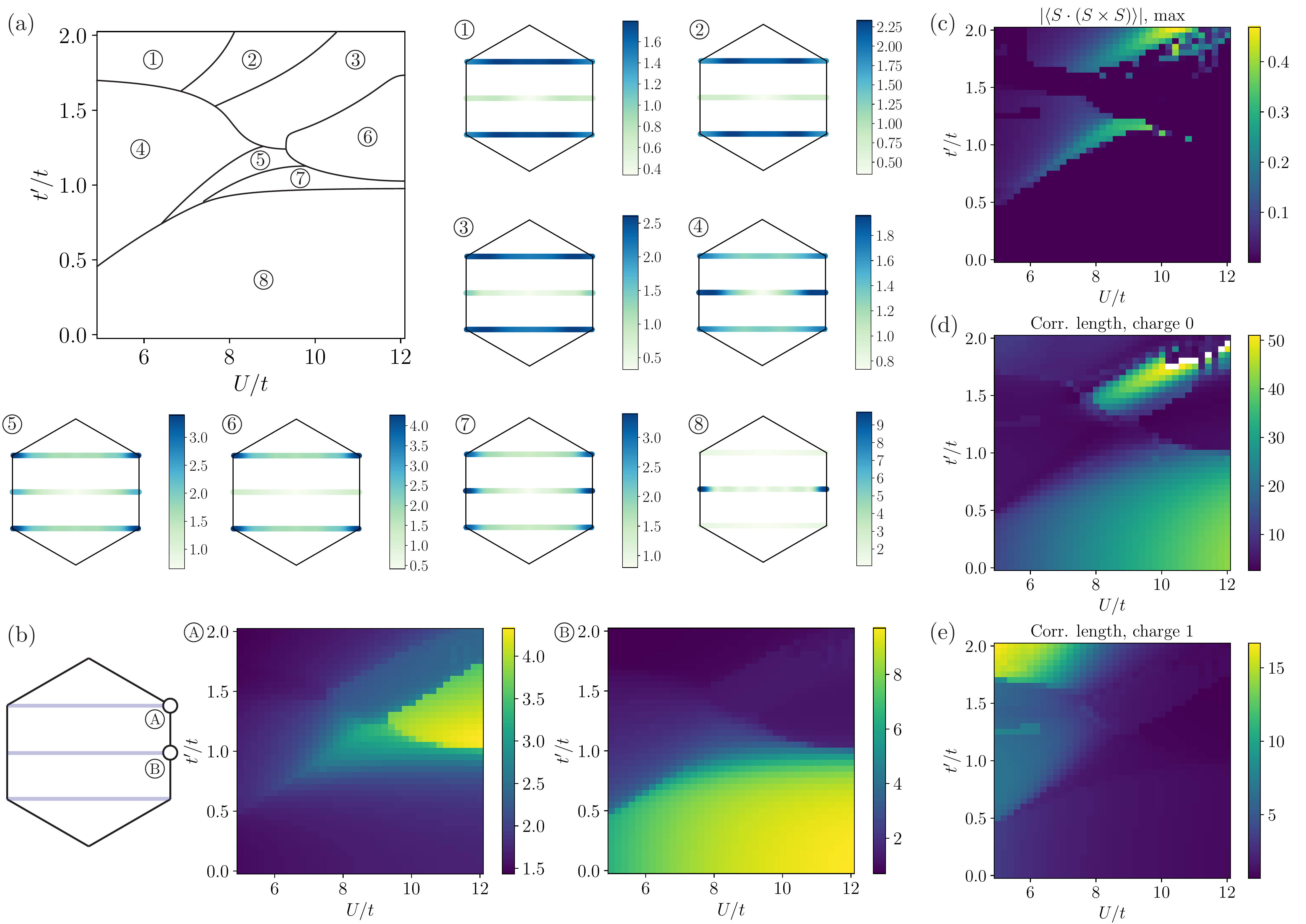}
\caption{Data for YC3 cylinder with symmetric anisotropy, with MPS bond dimension $\chi=4000$.  {\bf (a)} Spin structure factor, view 1.  We show the spin structure factor computed on allowed momentum cuts at one representative point in each of the eight phases we observe.  {\bf (b)} Spin structure factor, view 2.  We show, for the full range of $t'/t$ and $U/t$, the value of the spin structure factor at one $K$ point and one $M$ point in the Brillouin zone, labeled \Circled{A} and \Circled{B}, respectively.  The point \Circled{A} will have a peak for spiral order, while the point \Circled{B} will have a peak for N\'{e}el order.  {\bf (c)} Scalar chiral order parameter $\langle \mathbf{S}\cdot(\mathbf{S}\times\mathbf{S})\rangle$.  The magnitude and sign of the order parameter vary with a two-ring unit cell in phase \Circled{5} and a four-ring unit cell in phase \Circled{2}, so here we show for each parameter point the largest magnitude among all plaquettes.  In the Supplemental Material~\cite{SuppMat} we also show the smallest magnitude, which shows very similar behavior.  {\bf (d)} Correlation length for charge 0 excitations (i.e. spin excitations), computed using the MPS transfer matrix.  A large correlation length will be found if spin excitations are gapless.  {\bf (e)} Correlation length for charge 1 excitations.  A large correlation length is implied by gapless charge excitations.  The charge 2 correlation length can be found in the Supplemental Material, and is largest in phase \Circled{4}~\cite{SuppMat}.
\label{fig:YC3_sym_data}}
\end{figure*}

Data for the YC3 cylinder with symmetric anisotropy are shown in Fig.~\ref{fig:YC3_sym_data}, again calculated from MPS ground states with bond dimension $\chi=4000$.  Here all data are computed using a four ring unit cell with momentum 0 around the cylinder per unit cell.  The data were generated by adiabatically increasing and decreasing anisotropy from the isotropic line; part of the phase diagram also uses a dataset generated by adiabatically decreasing anisotropy starting from large $t'/t$.  Additional data are available in the Supplemental Material~\cite{SuppMat}.

Using these data, we identify the following phases:

\begin{itemize}[label={},leftmargin=*,listparindent=1em]
\item \emph{N\'{e}el order}: The bottom phase, labeled as \Circled{8} in Fig.~\ref{fig:YC3_sym_data}(a), is again square lattice N\'{e}el order, very similar to what was observed for the YC4 cylinder with symmetric anisotropy.

\item \emph{Spiral order}: The spiral-ordered phase, labeled by \Circled{6} in Fig.~\ref{fig:YC3_sym_data}(a), is again identifiable from the spin structure factor.  Here the allowed momentum cuts intersect the $K$ points, and this phase has peaks in the structure factor at those points, as expected for spiral (in particular, 120$^\circ$) order.  Surprisingly, this phase does not include the isotropic line $t'/t=1$, at least with the range of $U$ considered here, even though that is where the 120$^\circ$ order is expected in two dimensions.  This is a sign of the one-dimensionality of this small-circumference cylinder: coupling around the cylinder circumference apparently introduces some ``effective anisotropy'' so that $t'/t\approx 1.2$ acts like the isotropic line.

\item \emph{Chiral spin liquid}: The CSL, labeled by \Circled{5} in Fig.~\ref{fig:YC3_sym_data}(a), is again clearly identified by the scalar chiral order parameter, Fig.~\ref{fig:YC3_sym_data}(c), roughly corresponding to the green region nearer to the isotropic line.  The chiral order parameter here varies in sign and magnitude with a two-ring unit cell; the exact pattern is shown in the Supplemental Material~\cite{SuppMat}.  We remark that in our previous work~\cite{Szasz2020}, we reported the variation in magnitude (Fig.~11(a)), but did not note the varying sign; the data are in agreement.  The appearance of the CSL with an alternating sign for the scalar chirality on the YC3 cylinder, albeit with a different pattern, was recently also reported in Ref.~\cite{Zhu2020}.  Finally, we note that in our work on the isotropic model, we observed the CSL to be shifted to lower $U/t$ relative to the same phase in YC4 and YC6, but our new data reveal that to be an artifact of the aforementioned effective anisotropy---in fact, the phase appears at a similar range of $U/t$, just shifted upwards in $t'/t$.

\item \emph{Metal} or \emph{Luther-Emery liquid}: This phase, labeled by \Circled{4} in Fig.~\ref{fig:YC3_sym_data}(a), is identified by a large correlation length for both charge 1 [Fig.~\ref{fig:YC3_sym_data}(e)] and charge 2 (Supplemental Material~\cite{SuppMat}) excitations.  Note that charge 0 excitations being apparently gapped is only because the scale for Fig.~\ref{fig:YC3_sym_data}(d) is much larger than for (e); the same data as in (d) but with a cutoff at a correlation length of 12 are shown in the Supplemental Material\cite{SuppMat}.

\item \emph{1D Metal}: As for the YC4 cylinder with symmetric anisotropy, the upper left phase, labeled in Fig.~\ref{fig:YC3_sym_data}(a) by \Circled{1}, is characterized by gapless charge and an open Fermi surface.  ($k$-space occupation numbers are shown in the Supplemental Material~\cite{SuppMat}.)

\item \emph{Alternating chiral}: Again like the YC4 symmetric case, the next phase to the right, labeled by \Circled{2} in Fig.~\ref{fig:YC3_sym_data}(a), spontaneously breaks time-reversal symmetry and has nonzero local charge currents.  Here the scalar chiral order parameter, and the local currents, vary with a four-ring unit cell---the precise pattern is illustrated in the Supplemental Material~\cite{SuppMat}.  We note that since the calculations were performed with a four-ring unit cell and this phase is found to break translation symmetry with precisely the full four-ring unit cell, it may be that the true ground state has a larger unit cell or would have some noncommensurate pattern.

\item \emph{Spin liquid}: The upper right phase, labeled by \Circled{3} in Fig.~\ref{fig:YC3_sym_data}(a), appears to be a gapless spin liquid, with no magnetic order and a large spin correlation length.  To further understand this phase, we perform spin flux insertion for one representative point, finding the rather surprising behavior that there is an isolated gapless point at 0 flux as well as a large gapless region from roughly $\pi$ to $3\pi$ flux; see the Supplemental Material~\cite{SuppMat}.  This does not clearly correspond to a spin liquid state expected in two dimensions, in particular the 1D spin liquid found on the YC4 cylinder with symmetric anisotropy.

We also briefly note that there is some strange behavior visible in the very upper right in Figures \ref{fig:YC3_sym_data}(c) and (d); the simulation did not converge well here, presumably because the bond dimension is too small to well-approximate the gapless state.

\item \emph{Spiral 2}: Finally, the phase labeled by \Circled{7} in Fig.~\ref{fig:YC3_sym_data}(a) has a spin structure factor with roughly equal magnitude at the $K$ points and at the $k_y=0$ $M$ point, indicating some character of both 120$^\circ$ and N\'{e}el order.  This structure factor corresponds to real-space correlations (see the Supplemental Material~\cite{SuppMat}) that are antiferromagnetic along the cylinder and nearly 0 around the cylinder, which looks like an effective one-dimensional ordering; however, a quasi-one-dimensional state near the isotropic line would be quite surprising, so it may be better to interpret the minimal correlations around the cylinder as the way a more general two-dimensional order happens to manifest with this cylinder circumference.  

Then the natural possibility is another spiral phase.  In two dimensions, for $t'/t$ between 1 and $1/\sqrt{2}$, classically a spiral order with peak between the $K$ and $M$ points would be expected.  Given that such peak locations do not lie on the allowed momentum cuts, this state could manifest with smaller peaks at both high-symmetry points, which is what we observe in this phase.

\end{itemize}


\subsection{YC6, symmetric anisotropy}

\begin{figure*}
\includegraphics[width = \textwidth]{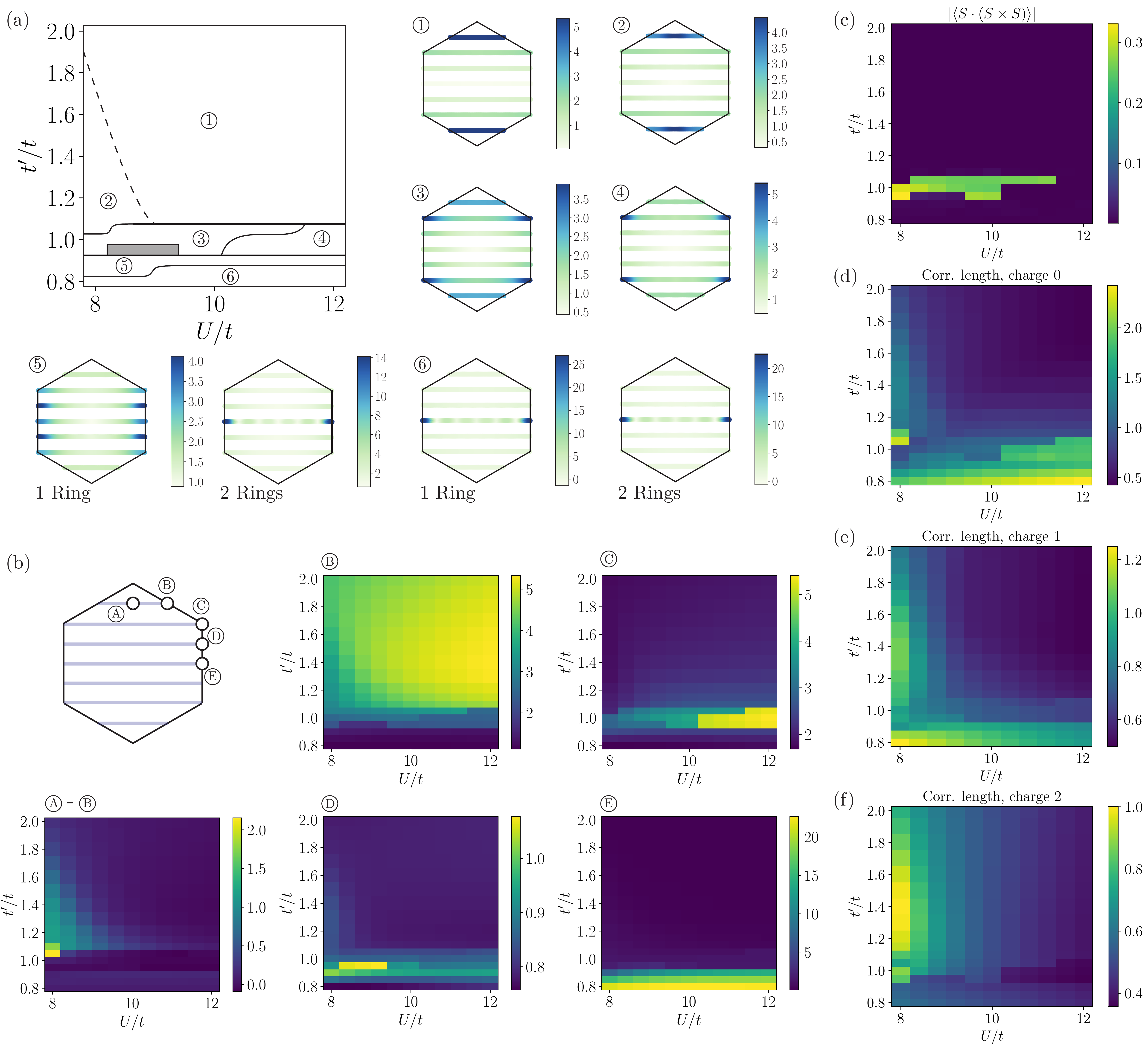}
\caption{Data for YC6 cylinder with symmetric anisotropy, with MPS bond dimension $\chi=8000$.  {\bf (a)} Spin structure factor, view 1.  We show the spin structure factor computed at specific points in the phase diagram, at least one point per phase.  The dashed line between \Circled{1} and \Circled{2} may correspond (roughly) to the location of a phase transition, though it is difficult to tell from our data.  For the phase labeled \Circled{5}, the states found with a one ring unit cell and with a two ring unit cell are qualitatively different and have a very small difference in energy, so we show the spin structure factor found with each setup.  We also show the spin structure factor for both setups for the phase labeled \Circled{6} to emphasize that here the two agree.  {\bf (b)} Spin structure factor, view 2.  We show, for the full range of $t'/t$ and $U/t$, the value of the spin structure factor at several important points in the Brillouin zone, namely at the right edge on each allowed momentum cut.  We also show the difference in height between points \Circled{A} and \Circled{B} to show that below the dashed line there is a small peak at \Circled{A}.  {\bf (c)} Scalar chiral order parameter $\langle \mathbf{S}\cdot(\mathbf{S}\times\mathbf{S})\rangle$.  {\bf (d)} Correlation length for charge 0 excitations (i.e. spin excitations), computed using the MPS transfer matrix. {\bf (e)} Correlation length for charge 1 excitations.  {\bf (f)} Correlation length for charge 2 excitations.
\label{fig:YC6_sym_data}}
\end{figure*}

Data for the YC6 cylinder with symmetric anisotropy are shown in Fig.~\ref{fig:YC6_sym_data}, computed from MPS ground states with bond dimension $\chi=8000$.  For YC6, we use a two ring unit cell with momentum 0 around the cylinder, as well as a one ring unit cell initialized both with momentum 0 and with momentum $\pi$ per ring.  Note that although this bond dimension is larger than that used for the smaller cylinders, the needed bond dimension for a given level of precision scales roughly exponentially in the circumference, so these results are relatively less well converged than for the smaller cylinders.  In the large $t'/t$ limit, we have checked several points using a larger bond dimension of $16000$ and found no qualitative difference.  Note that due to increased computational cost compared with smaller cylinders, we use a lower resolution in parameter space and also restrict the range of $U/t$ and $t'/t$ to the region where the results on the smaller cylinders show the strongest disagreement.  Some additional data not included in Fig.~\ref{fig:YC6_sym_data} (entanglement spectra and real-space spin correlations) can be found in the Supplemental Material~\cite{SuppMat}.

Using these data, we identify the following phases:

\begin{itemize}[label={},leftmargin=*,listparindent=1em]
\item \emph{N\'{e}el order}: The bottom phase, labeled as \Circled{6} in Fig.~\ref{fig:YC6_sym_data}(a), is again square lattice N\'{e}el order, very similar to what was observed for the other two cylinders with symmetric anisotropy.

\item \emph{Spiral order}: The spiral-ordered phase, labeled by \Circled{4} in Fig.~\ref{fig:YC6_sym_data}(a), is identifiable by clear peaks in the spin structure factor at the $K$ points as expected for 120$^\circ$ order. 

\item \emph{Chiral spin liquid}: The CSL, labeled by \Circled{3} in Fig.~\ref{fig:YC6_sym_data}(a), is again clearly identified by the scalar chiral order parameter, Fig.~\ref{fig:YC6_sym_data}(c).  The region shaded gray in Fig.~\ref{fig:YC6_sym_data}(a) may also belong to the CSL.  On this $t'/t=0.95$ line, we find nonzero scalar chirality with momentum 0 per ring and zero scalar chirality with momentum $\pi$.  The 0-momentum state is lower in energy for both higher and lower $U$, with the $\pi$-momentum state being lower in energy for the three values of $U/t$ that comprise the shaded region.  On the whole line, the difference in energy is less than the error due to truncation of the MPS bond dimension while running DMRG, especially so in the shaded region.  (In contrast, for $t'/t=1.0$ and 1.05, both momentum sectors show nonzero chirality, so the phase identification there is quite clear.)  If the gray shaded region does not belong to the CSL, then it, or indeed the whole $t'/t=0.95$ line, could belong instead to the spiral order.

\item \emph{N\'{e}el/Spiral order}: This region of parameter space, labeled by \Circled{5} in Fig.~\ref{fig:YC6_sym_data}(a), may belong to the N\'{e}el phase or to the spiral phase.  When simulations are performed with a two ring unit cell, allowing translation symmetry breaking, the ground state has clear N\'{e}el order.  With a one ring unit cell the ground state instead has a spin structure factor with peaks at the edge of the Brillouin zone on the $k_y=\pm\pi/2$ lines, corresponding to spiral order; in particular, this appears to be the same spiral order as in the ``Spiral 2'' phase for YC3.  While the two ring state has slightly lower energy, the difference is small, and with larger bond dimensions the order might reverse since small bond dimensions favor symmetry breaking to reduce entanglement.  Thus whether this region actually belongs to the N\'{e}el phase or the spiral phase is uncertain.  

\item \emph{1D spin liquid}: Much of the upper portion of the phase diagram appears to be a single phase, labeled by \Circled{1} in Fig.~\ref{fig:YC6_sym_data}(a).  We tentatively identify this phase as the 1D spin liquid, because the spin structure factor, correlation lengths, and entanglement spectrum (shown in the Supplemental Material~\cite{SuppMat}), are much the same as for the 1D spin liquid phase on the YC4 cylinder with symmetric anisotropy, and the spin structure factor again matches what was reported in Refs.~\cite{Heidarian2009,Tocchio2014}.  However, for this phase we do not have flux insertion data to fully confirm this identification.

\item \emph{Possible additional phase}: Finally, the region below the dashed line in Fig.~\ref{fig:YC6_sym_data}(a), labeled \Circled{2}, may be an additional phase.  All computed quantities change smoothly from here into phase \Circled{1}, but there are notable qualitative differences, including much longer correlation lengths.  Most notably, the spin structure factor shows significantly different behavior, with a peak at $(0,\pi)$, whereas in region \Circled{1} there is a uniform peak along the $k_y=\pi$ line.  The existence of a peak corresponds to extended correlations along the cylinder, while the uniform maximum along a line indicates close to zero correlation along the cylinder; see the real-space correlations in the Supplemental Material~\cite{SuppMat}.

\end{itemize}


\section{Discussion\label{sec:discuss}}

We have used density matrix renormalization group simulations to map out the phase diagram of the triangular lattice Hubbard model with anisotropic hopping---the bonds along one of the three orientations in the lattice have hopping strength $t'$, while bonds in the other two orientations have strength $t$.  We studied the phase diagram as a function of both interaction strength $U/t$ and anisotropy $t'/t$, from the square lattice limit at $t'=0$ to the weakly coupled chain limit of large $t'$.  Using four distinct cylinder geometries, namely circumferences of three, four, and six sites with the distinct bond around the cylinder (symmetric anisotropy) and circumference four with the distinct bond on a diagonal (asymmetric anisotropy), we find a large variety of phases.

For the three larger cylinders, right around the isotropic line $t'/t=1$ we observe the same phases as in our previous work~\cite{Szasz2020}, namely an apparently metallic phase that is likely a Luther-Emery liquid, a chiral spin liquid, and spiral magnetic order, exemplified by the 120$^\circ$ order expected at high $U$ exactly at $t'/t=1$.  On the smallest cylinder, with circumference three, these phases are all shifted upwards, being centered instead on $t'/t\approx 1.2$; there is apparently some ``effective anisotropy'' as a result of the small cylinder circumference. Conversely, the consistent behavior around the isotropic line on the three larger cylinders, in particular that the CSL is centered around and spiral order is strongest near $t'/t = 1$, strongly indicates that these cylinders are large enough to give reliable results for the two-dimensional limit of the isotropic Hubbard model.

Towards the square lattice limit, with smaller $t'$, the results are again quite consistent among the four different cylinders: the square lattice N\'{e}el magnetic order takes up a large portion of parameter space, already being stabilized with around 20\% anisotropy.  We also observe on the circumference four cylinder with asymmetric anisotropy a small phase that appears to be a gapless spin liquid, and for circumference three and possibly circumference six, we find a phase between the spiral and N\'{e}el orders at high $U$ that appears to be a distinct magnetically ordered state, but in fact should also belong to the spiral phase for the two-dimensional model.

For large $t'$, we observe a tremendous variety of phases, with little agreement among the different cylinders.  One phase that is consistently present is an apparently metallic state, with large charge correlation length, at low $U$ and large $t'$, which based on occupation numbers in the Brillouin zone appears to have an open Fermi surface, consistent with the $U=0$ state for $t'/t\gtrsim 1.64$. 
While this state is clearly distinct from the metal/LEL for the YC3 and YC4 cylinders with symmetric anisotropy, it is not clear whether this will still be true in the two-dimensional limit---on these cylinders the Fermi surface becomes gapped on one of a small number of allowed momentum cuts, so it is not surprising that this produces a dramatic change in the ground state.
Beyond this, we find various spin liquid candidates, including an apparently gapless spin liquid just above the CSL as well as a gapped nonmagnetic state in the same place in the phase diagram on a different cylinder, and a ``1D spin liquid'' at large $t'$ and $U$.  We also observe collinear magnetic order and phases with an alternating scalar chiral order parameter and accompanying local currents.  Notably, the YC6 cylinder shows fewer phases than the smaller cylinders, even accounting for the fact that we considered a smaller portion of parameter space.  That the 1D spin liquid is the phase that survives at large $U$ and large $t'$ in this, the largest and most two-dimensional of the cylinders we study, suggests that this phase is perhaps the strongest candidate for this parameter regime on the full two-dimensional lattice.

Comparing these results with those of past theoretical works using such techniques as variational Monte Carlo and dynamical mean field theory, summarized in Sec.~\ref{sec:theory} above, the general trends are similar: the large and very stable N\'{e}el phase for $t'/t<1$ certainly comports with past works, while many of the phases we find in the $t'/t > 1$ regime, for example the 1D spin liquid and collinear order, have been predicted before.  We do also find phases that have never been predicted before, namely the phases in circumference three and four cylinders with alternating chiral order parameters and local currents; however, these phases do not seem likely to survive in the two-dimensional limit, especially since we already find no indication of this behavior with circumference six.  One phase found in some past works of which we find no indication is a superconducting state between the metal/Luther-Emery liquid and N\'{e}el phases.  

We can now consider the implications of our work for understanding experiments on nearly isotropic triangular lattice spin liquid candidates like $\kappa$-(BEDT-TTF)$_2$Cu$_2$(CN)$_3$.  These materials do generally have small hopping/spin-exchange anisotropies, on the order of 10 or 20\%, so in studying only the isotropic line, it is not clear whether a predicted spin liquid phase is actually a strong candidate for the nature of the experimentally observed nonmagnetic insulators.  We find here that the chiral spin liquid from the isotropic line is stable to around 10\% anisotropy.  That it is in fact stable with some anisotropy rather than existing purely on the isotropic line means that it does remain a viable candidate state worth taking into consideration, but on the other hand it is also sufficiently unstable that if the true anisotropy of spin liquid candidate materials turns out to be on the larger side of the various estimates, our simulations would no longer indicate the CSL as the likely ground state.  The other relevant finding for nearly isotropic candidate materials is that there seem to emerge gapless spin liquids for some cylinder geometries both above and below the CSL.  While it is not clear whether these states indeed survive to the two-dimensional limit, their presence on some cylinders at least shows they are relatively low energy states in the Hubbard model. Therefore, they could be stabilized by, for example, longer range interactions in real materials, implying that the Hubbard model does suggest gapless spin liquids as reasonable candidates for the states observed in experiments.  In principle, our results could also speak to the nature of the putative spin liquid states observed in Cs$_2$CuCl$_{4-x}$Br$_x$ compounds, which are less isotropic with $t'/t>1$, but the large variety of phases we find in the $t'/t>1$ limit makes it difficult to make a prediction with confidence. Furthermore, the large Hubbard-$U$ in these materials~\cite{Foyevtsova2011} means that the less computationally expensive Heisenberg model is likely the preferrable choice for numerical simulations.

Finally, we wish to highlight some of the most important open questions following our work.  Of course it continues to be the case that there are a great many possible phases in the $t'/t>1$ limit, and which ones will survive to the full two-dimensional model remains unclear.  For future DMRG studies of this problem, this can be addressed both by pushing to higher bond dimensions and larger cylinders as computational resources improve, and in the nearer term by using a wider variety of cylinder geometries such as XC cylinders where one bond of the triangular lattice runs along the length of the cylinder, which will give further intuition about which phases tend to be stable on a large variety of cylinders and thus might appear in two dimensions.  Moving beyond DMRG, a promising approach is to study the model directly in the two-dimensional limit using, for example, projected entangled pair states~\cite{Verstraete2004,Corboz2010}.

There is also still further work to be done to better understand the phase diagrams on these finite-circumference cylinders.  For example, the gapless spin liquid phase above the CSL on the circumference four cylinder with symmetric anisotropy appears to be gapless only near zero spin-flux, indicating a possible Dirac or otherwise nodal gapless spin liquid, but because matrix product states cannot easily describe gapless states, significant computational effort to study scaling with very large bond dimensions would be needed to conclusively identify this state.  This possible Dirac spin liquid and many of the other phases we find for $t'/t>1$ are interesting in their own right, even if they may not appear in the full two-dimensional model, so these further calculations are well worth pursuing.


\begin{acknowledgments}

We would like to thank Yin-Chen He and Michael Zaletel for helpful discussions.  We have also used the TenPy tensor network library~\cite{Kjall2013}, which includes contributions from Zaletel, Roger Mong, Frank Pollmann, and others.  Numerical computations were primarily performed using the Lawrencium cluster at Lawrence Berkeley National Laboratory; some computations were also performed using the Symmetry cluster at Perimeter Institute.  
J.M. received funding through DFG research fellowship No.\ MO 3278/1-1 and TIMES at Lawrence Berkeley National Laboratory supported by the U.S. Department of Energy, Office of Basic Energy Sciences, Division of Materials Sciences and Engineering, under Contract No.\ DE-AC02-76SF00515.
Research at Perimeter Institute is supported in part by the Government of Canada through the Department of Innovation, Science and Economic Development Canada and by the Province of Ontario through the Ministry of Colleges and Universities.

\end{acknowledgments}


%

%
\bigskip

\onecolumngrid
\newpage



\section*{Supplementary material}

# Phase diagram of the anisotropic triangular lattice Hubbard model

Aaron Szasz[1, 2, *] and Johannes Motruk[3, 2]

[1]*Perimeter Institute for Theoretical Physics, Waterloo, Ontario N2L 2Y5, Canada*

[2]*Materials Sciences Division, Lawrence Berkeley National Laboratory, Berkeley, California 94720, USA*

[3]*Department of Physics, University of California, Berkeley, California 94720, USA*

(Dated: May 12, 2021)

## I. MAGNETIC ORDERS

Here we lay out the different types of long-range magnetic orders that we might expect to appear on the triangular lattice. For each one, we compute the spin structure factor

$$S(\mathbf{q}) = \frac{1}{N} \sum_{ij} \langle \mathbf{S}_i \cdot \mathbf{S}_j \rangle e^{i\mathbf{q} \cdot \mathbf{r}_{ij}}, \tag{1}$$

where $N$ is the total number of sites in the lattice, and then discuss when such an ordering might appear for a full two-dimensional triangular lattice antiferromagnet, as well as how it could manifest in each of the four specific cylinder geometries we use: YC4 with symmetric and asymmetric anisotropy, and YC3 and YC6 with symmetric anisotropy.

For concreteness, we begin by specifying a particular setup for our lattice vectors and reciprocal lattice vectors, as shown in Figure S1, namely

$$\mathbf{a} = (0, 1), \quad \mathbf{b} = \left( \frac{\sqrt{3}}{2}, \frac{1}{2} \right) \tag{2a}$$

$$\mathbf{k}_a = \frac{4\pi}{\sqrt{3}} \left( -\frac{1}{2}, \frac{\sqrt{3}}{2} \right), \quad \mathbf{k}_b = \frac{4\pi}{\sqrt{3}} (1, 0). \tag{2b}$$

Recall that these satisfy

$$\mathbf{a} \cdot \mathbf{k}_a = \mathbf{b} \cdot \mathbf{k_b} = 2\pi, \quad \mathbf{a} \cdot \mathbf{k}_b = \mathbf{b} \cdot \mathbf{k_a} = 0. \tag{3}$$

In terms of these vectors, we will write the real-space separation between sites as

$$\mathbf{r} = n_a \mathbf{a} + n_b \mathbf{b} \tag{4}$$

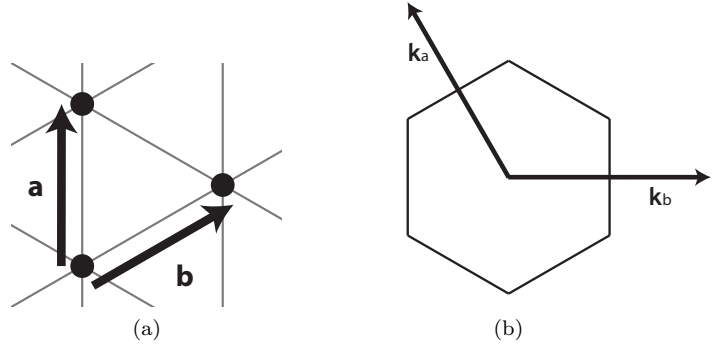

FIG. S1. (a) Lattice vectors $\mathbf{a}$ and $\mathbf{b}$ on the triangular lattice. (b) Corresponding reciprocal lattice vectors and the first Brillouin zone.

———————

* aszasz@perimeterinstitute.ca

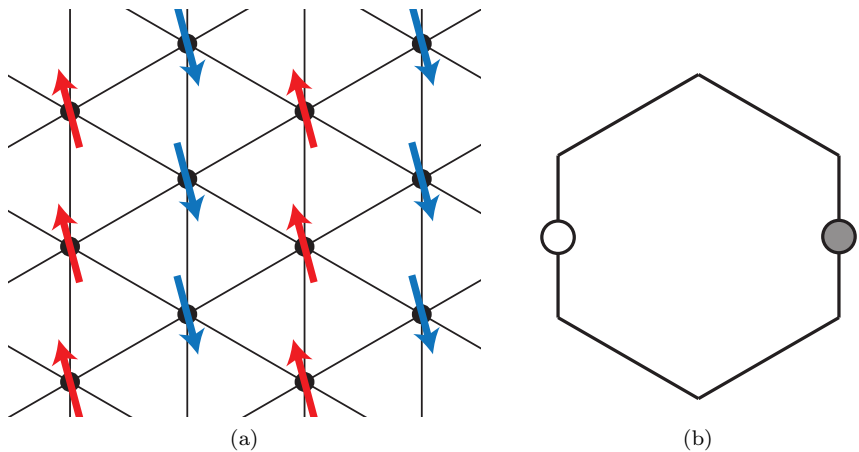

FIG. S2. (a) Real-space picture of magnetic ordering which is ferromagnetic along one direction in the triangular lattice (in this case the vertical direction, along **a**). Color-coding is provided for clarity but is not physically meaningful. Note that the precise direction of the aligned spins is not important, only the relative orientations of the spins; in other words, there is a global $SO(3)$ symmetry, and we show the pattern for an arbitrary choice of direction. (b) Corresponding spin structure factor. The filled circle denotes the location of the diverging peak at the $M$ point, while the unfilled circle shows the paired $M$ point, which is equivalent under translation by a reciprocal lattice vector.

where $n_a$ and $n_b$ are integers, and the momentum as

$$\mathbf{q} = m_a \mathbf{k}_a + m_b \mathbf{k}_b \tag{5}$$

where $m_a$ and $m_b$ are each in $(-0.5, 0.5]$. Note that $\mathbf{q} \cdot \mathbf{r}$ then becomes $2\pi(m_a n_a + m_b n_b)$. We further assume that the lattice consists of $N_a \times N_b$ unit cells, with $N = N_a N_b$, letting both quantities diverge to find the true spin structure factor, $S(m_a, m_b)$.

### A. One direction ferromagnetic, two directions antiferromagnetic

*Abstract picture*: The first type of ordering we consider is shown in Figure S2(a). Evidently, along one lattice vector the spins are ferromagnetically aligned, while along the two other bonds of the triangular lattice they are antiferromagnetically ordered. In terms of the lattice vectors specified above, we see that $\langle \mathbf{S}_i \cdot \mathbf{S}_j \rangle = (-1)^{n_b} = e^{i\pi n_b}$. Then

$$S(m_a, m_b) = \sum_{n_a, n_b} e^{i\pi n_b} e^{i2\pi(m_a n_a + m_b n_b)} \tag{6a}$$

$$= \left( \sum_{n_a} e^{i2\pi m_a n_a} \right) \left( \sum_{n_b} e^{i2\pi(m_b + 1/2)n_a} \right) \tag{6b}$$

$$= N \delta_{m_a, 0} \delta_{m_b, 1/2} \tag{6c}$$

In other words, the structure factor is exactly 0 except at the points $\pm \mathbf{k}_b/2$, where it diverges linearly in the number of lattice sites; these are two of the so-called $M$ points of the Brillouin zone. (Note that above we specified $\delta_{m_b, 1/2}$, which specifies only $\mathbf{k}_b/2$, but this is equivalent to $-\mathbf{k}_b/2$ since they are connected by a reciprocal lattice vector.) This structure factor is shown in Figure S2(b).

*Anisotropic triangular lattice*: There are two main scenarios in which this order might appear on a 2D anisotropic triangular lattice antiferromagnet. The first is the limit in which one bond is much weaker than the remaining two, shown in Figure S3(a), in which case this ordering is actually just the square lattice Néel antiferromagnet, and the ferromagnetic direction is oriented along the weak bond.

The second is the limit where one bond is much stronger than the remaining two, shown in Figure S3(b), in which case the magnetic order is referred to as "collinear." In that case, one antiferromagnetic direction follows the strong bond, and among the remaining two directions there is no preference for which is ferrogmagnetic and which is antiferromagnetic. This symmetry might be spontaneously broken, in which case we would expect to see a diverging structure factor at just one pair of $M$ points (on the edges of the Brillouin zone that are parallel to the ferromagnetic

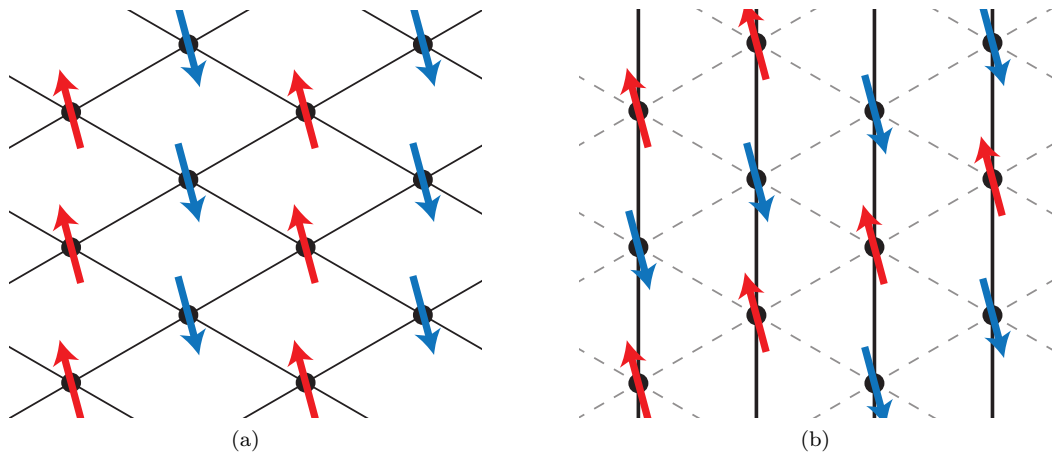

FIG. S3. (a) The one direction ferromagnetic, two directions antiferromagnetic order is expected in the square lattice limit of the anisotropic triangular lattice, where one bond is much weaker than the other two. As shown here, the ferromagnetic direction is along the weak bond. In this context, the magnetic ordering is just the square lattice Néel aniferromagnetic order. (b) It can also occur in the limit of one strong bond (solid line) in which case one of the antiferromagnetic directions is along that bond. The remaining two directions (dashed lines) are in principle equivalent, so on the full lattice, which one is ferromagnetic is random if the symmetry is spontaneously broken (shown here), or the symmetry could be preserved. In this context, the ordering is referred to as "collinear."

direction), or the symmetry might be preserved, in which case the structure factor would have peaks at two pair of $M$ points, only leaving empty the pair on the edges of the Brillouin zone that are parallel to the strong bond direction.

*In our simulations*: All six $M$ points are included on the allowed momentum cuts for YC4 and YC6, so any orientation of this magnetic order is allowed on those cylinders. For YC3, only the $M$ points with $k_y = 0$ are included, so we expect that we will only find this ordering if the ferromagnetic direction is around the cylinder circumference.

When the ordering arises in the square lattice limit of one very weak bond, in the case of symmetric anisotropy the ferromagnetic direction will indeed be around the cylinder, so we expect to observe the square lattice Néel order for all three cylinder circumferences, with structure factor peaks at $\pm k_b/2$. For the YC4 asymmetric anisotropy case, the order is still allowed and we again expect to observe it, but with the structure factor peaks instead at $\pm(k_a + k_b)/2$.

When the ordering instead arises in the limit of one very strong bond, with symmetric anisotropy we would expect to see the ferromagnetic ordering along one of the diagonal directions, so the structure factor would have peaks either at $\pm(k_a + k_b)/2$ or at $\pm k_a/2$, or at both sets of $M$ points if the symmetry between the two possibilities is not broken. In this case the model itself does not break the symmetry, but the simulation does allow it to be spontaneously broken.

With the asymmetric anisotropy, the strong bond is along $\mathbf{b}$, so the structure factor would have peaks at $\pm k_b/2$ if the ferromagnetic direction is around the cylinder, or at $\pm(k_a + k_b)/2$ if the ferromagnetic direction is along the other diagonal ($\mathbf{b} - \mathbf{a}$). In this case, the model itself breaks the symmetry between these two possibilities, so we expect at least to observe that one pair of peaks is stronger than the other.

## B. One direction antiferromagnetic, others zero

*Abstract picture*: Another possible ordering is antiferromagnetic along one direction and essentially zero in the remaining directions. We note that this is not true long-range order—as we will see shortly, there is only a sub-extensive divergence in the spin structure factor in the thermodynamic limit. However, there is long-range order along one direction in the lattice, which makes this an important case to consider when working with cylinders. This order is shown in Figure S4(a). Note that we draw a technically incorrect figure which shows zero spin off of one particular line in the lattice; the figure can be viewed as giving directions for computing $\langle \mathbf{S}_i \cdot \mathbf{S}_j \rangle$ rather than as showing $\langle \mathbf{S} \rangle$ on each site.

The real-space spin correlations are given by $\langle \mathbf{S}_i \cdot \mathbf{S}_j \rangle = e^{i\pi n_a}\delta_{n_b}$, which gives

$$S(m_a, m_b) = N_a \delta_{m_a, 1/2}. \tag{7}$$

As promised, this has a sub-extensive divergence in the thermodynamic limit, but in numerics on finite-circumference cylinders it is still very much observable. The resulting spin structure factor is shown in Figure S4(b).

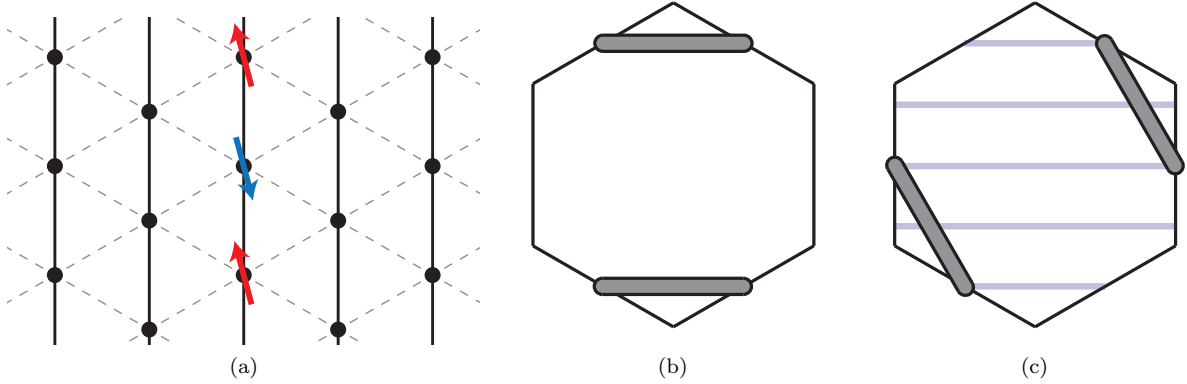

FIG. S4. (a) Spin ordering with antiferromagnetic correlations along one direction and zero correlation along other directions. Note that because there is no two-dimensional long range order, this figure does not show the magnetic moment on each site but rather the antiferromagnetically-ordered moments along one line and zero moment everywhere else, indicating the lack of correlation. (b) (Sub-extensive) peaks in the spin structure factor are flat lines between $M$ points. (c) For the YC4 cylinder with asymmetric anisotropy, the structure factor peak lines intersect all allowed momentum cuts as shown here.

*Anisotropic triangular lattice*: This type of ordering could appear when one bond is far stronger than the other two; indeed, in the limit where two bonds go precisely to 0, this is expected for the short-range spin ordering, though the long-range one-dimensional order is not possible since it would have to break a continuous symmetry.

*In our simulations*: With the symmetric anisotropy, the strong bond would be around the cylinder, and thus the structure factor peak for this order would lie precisely along one of the allowed momentum cuts for both YC4 and YC6, meaning that this kind of order could be observed. Also note that although spontaneously breaking a continuous symmetry to get long-range one-dimensional order is not possible, in this case the ordering would be just around the finite cylinder circumference so this kind of spin ordering is not blocked. For the YC3 cylinder with symmetric anisotropy, this ordering is geometrically frustrated; i.e. the structure factor peak is not on an allowed momentum cut; we therefore don't expect to observe it.

With the asymmetric anisotropy on the YC4 cylinder, the lines where the structure factor is peaked intersect all allowed momentum cuts, as shown in Figure S4(c), and the height of the structure factor would be equal at all intersections. In practice, it would be unlikely to observe precisely this behavior, due to the broken symmetry in the model: there is likely to be at least some residual correlation around the relatively small cylinder circumference, which would modify the structure factor to give some character of the collinear phase, making the peak higher near one pair of $M$ points than the other.

## C. Spiral order

*Abstract picture*: The last main type of spin ordering that we expect is the spiral order, of which the 120° phase is the most famous example. We first show this special case before considering the more general order.

The 120° order is shown in Figure S5(a). From the figure, we can see that $\langle \mathbf{S}_i \cdot \mathbf{S}_j \rangle = \cos(2\pi(n_a - n_b)/3)$. Then

$$S(m_a, m_b) = \sum_{n_a, n_b} \cos(2\pi(n_a - n_b)/3) e^{i2\pi(m_a n_a + m_b n_b)} \tag{8a}$$

$$= \frac{1}{2} \left( \sum_{n_a} e^{i2\pi(m_a + 1/3)n_a} \sum_{n_b} e^{i2\pi(m_b - 1/3)n_b} + \text{H.c.} \right) \tag{8b}$$

$$= \frac{N}{2} \left( \delta_{m_a, -1/3} \delta_{m_b, 1/3} + \delta_{m_a, 1/3} \delta_{m_b, -1/3} \right). \tag{8c}$$

This structure factor has divergences precisely at the $K$ and $K'$ points, the corners of the Brillouin zone:

$$K = \mathbf{k}_a/3 - \mathbf{k}_b/3 = \frac{2\pi}{3} \left( -\sqrt{3}, 1 \right) \tag{9a}$$

$$K' = \mathbf{k}_b/3 - \mathbf{k}_a/3 = \frac{2\pi}{3} \left( \sqrt{3}, -1 \right), \tag{9b}$$

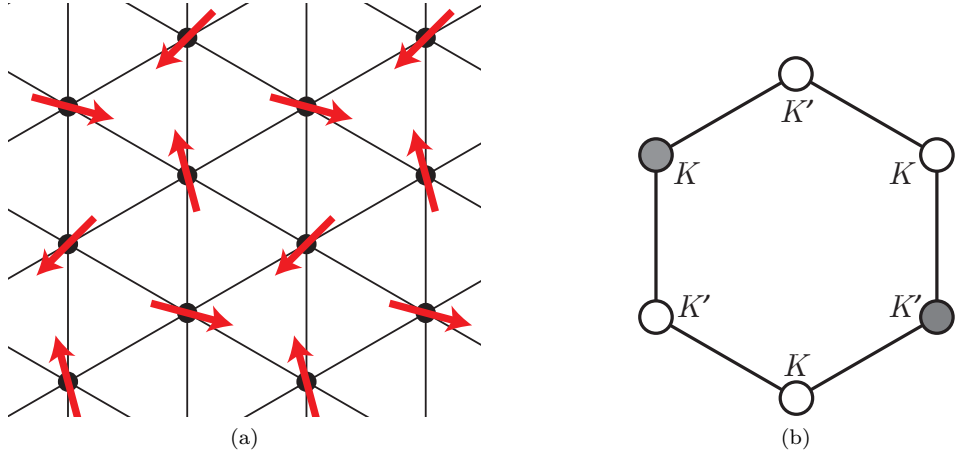

FIG. S5. (a) 120° magnetic order in real space, and (b) the corresponding spin structure factor. Filled circles indicate the $K$ and $K'$ points as defined in equation (9), and open circles indicate the points that are equivalent under translation by a reciprocal lattice vector.

shown in Figure S5(b). Note that these two points are inequivalent under translation by a reciprocal lattice vector, but each of the remaining four corners is equivalent to one of these two.

Now we consider a more general spiral order, in which the spin direction rotates by angle $\theta$ when moving one lattice spacing in direction $\mathbf{a}$ and by $\theta'$ when moving one lattice spacing in direction $\mathbf{b}$. This order is shown in Figure S6(a). The real-space spin correlations then satisfy $\langle \mathbf{S}_i \cdot \mathbf{S}_j \rangle = \cos(\theta n_a + \theta' n_b)$. This does indeed describe the spiral order in the case that $\theta = 2\pi/3$ and $\theta' = -2\pi/3$, and it also describes the one-FM, two-AFM order (Section I A), with the ferromagnetic direction along $\mathbf{a}$, when $\theta = \pi$ and $\theta' = 2\pi$, or along $\mathbf{b}$ for $\theta = 2\pi$ and $\theta' = \pi$. Following the same type of calculation as for the 120° special case, we see that

$$S(m_a, m_b) = \frac{N}{2} \left( \delta_{m_a, \theta/(2\pi)} \delta_{m_b, \theta'/(2\pi)} + \delta_{m_a, -\theta/(2\pi)} \delta_{m_b, -\theta'/(2\pi)} \right). \tag{10}$$

The diverging peaks in the structure factor are located at

$$K(\theta) = \frac{\theta}{2\pi} \mathbf{k}_a + \frac{\theta'}{2\pi} \mathbf{k}_b = \left( \frac{2\theta' - \theta}{\sqrt{3}}, \theta \right) \tag{11a}$$

$$K'(\theta) = -\frac{\theta}{2\pi} \mathbf{k}_a - \frac{\theta'}{2\pi} \mathbf{k}_b = -\left( \frac{2\theta' - \theta}{\sqrt{3}}, \theta \right). \tag{11b}$$

*Anisotropic triangular lattice*: The 120° order has been well-established [1–3] as the ground state at the isotropic point of the antiferromagnetic Heisenberg model and thus in the high-$U$ limit of the Hubbard model. Moving away from the isotropic point, a more general spiral order is expected. We can guess what kind of spiral order to expect based on a classical analysis of the anisotropic Heisenberg model; the result is that when the bond that is weaker or stronger than the other two lies along $\mathbf{b}$, the peak in the spin structure factor is expected to be at

$$\frac{q}{2\pi} \left( \mathbf{k}_b + \mathbf{k}_a \right) + \frac{q}{2\pi} \mathbf{k}_b = \frac{q}{2\pi} \left( \mathbf{k}_a + 2\mathbf{k}_b \right); \tag{12}$$

to derive this from the results following Equation (2) of reference 4, note that the lattice vectors used in the reference correspond to our $\mathbf{a}$ and $\mathbf{b} - \mathbf{a}$, for which the reciprocal lattice vectors are $\mathbf{k}_a + \mathbf{k}_b$ and $\mathbf{k}_b$. Likewise, when the distinct bond lies along $\mathbf{a}$, the lattice vectors of reference 4 are $\mathbf{k}_a - \mathbf{k}_b$ and $\mathbf{k}_b$, with reciprocal vectors $\mathbf{k}_a + \mathbf{k}_b$ and $\mathbf{k}_a$, so that the peak in the structure factor is at

$$\frac{q}{2\pi} \left( \mathbf{k}_b + \mathbf{k}_a \right) + \frac{q}{2\pi} \mathbf{k}_a = \frac{q}{2\pi} \left( 2\mathbf{k}_a + \mathbf{k}_b \right). \tag{13}$$

In either orientation, the classically predicted value of $q$ is $\pi$ when $t'/t < \sqrt{2}$, giving the Néel order, and $q = \arccos\left( -(t/t')^2/2 \right)$ for $t'/t > \sqrt{2}$ [4]; when $t' = t$, this gives $q = 2\pi/3$, and when $t'/t \to \infty$, it gives $q = \pi/2$.

Putting this in the language of $\theta$ and $\theta'$ used above, we see that for the distinct bond along $\mathbf{a}$, $\theta = 2\theta' = 2q$, whereas for the distinct bond along $\mathbf{b}$, $\theta' = 2\theta = 2q$. The locations of the peaks are illustrated, for the full classical

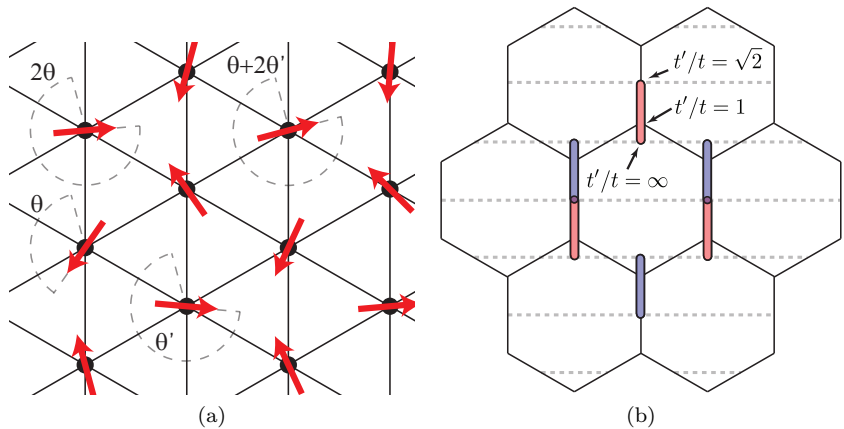

FIG. S6. (a) Real-space picture of more general spiral order. The local magnetic moment rotates by angle $\theta$ when moving one lattice site along **a** and by $\theta'$ along **b**; angles shown in the figure are with reference to the lower-left spin. (b) In the specific case of the anisotropic triangular lattice Heisenberg model, when the distinct bond is along **a**, classical spin-wave analysis gives $\theta = 2\theta'$ with $\theta' \in [\pi/2, \pi]$. When viewed purely in the first Brillouin zone, the resulting structure factor peak appears to jump discontinuously at $\theta' = 2\pi/3$, i.e. at the point with 120° order; to make clear that there is no physical discontinuity, we show the line of peak locations in a larger reciprocal lattice space. The regions indicated on the center vertical line are the ones from equations (11), and the other shaded regions indicate where the peaks would additionally appear in the first Brillouin zone. The red-shaded regions are inequivalent to the blue-shaded ones. Dashed lines are a guide to the eye to indicate $M$ points. We also indicate the specific peak locations expected from the classical analysis for several important values of $t'/t$: 1, $\sqrt{2}$, and $\infty$.

range $q \in [\pi/2, \pi]$, in Figure S6(b). The peaks move from the $M$ point for Néel order at $q = \pi$ along the edge of the Brillouin zone to the $K$ points at $q = 2\pi/3$, then into the Brillouin zone along the same straight line until reaching the center of the line directly between the other two $M$ points.

Apart from the location of the peaks, we can also consider whether long-range order is actually energetically favorable, or equivalently how much the classical ordering should be suppressed by quantum fluctuations. Linear spin wave analysis [4] shows that magnetic ordering is expected to be quite strong near the isotropic point, with the 120° order, and to vanish in both limits of $q \to \pi/2$ and $q \to \pi$.

*In our simulations*: Here the particular cylinder geometries play a very large role in allowing or disallowing certain orders. Beginning with the 120° order expected on the isotropic line, it will be allowed if the momentum cuts from the cylinder pass through the $K$ points, which is true for YC3 and YC6 but not for YC4. However, what we observe in practice is that something like this ordering does still occur for YC4: we see peaks in the structure factor at the points on the allowed momentum cuts that are closest to the $K$ points.

Moving away from the isotropic point, we would naively expect the magnetic order in the spiral phase to appear strongest when the ordering is allowed by the cylinder circumference, or equivalently when the peak in the structure factor lies on one of the allowed momentum cuts. This picture gives slightly different predictions for each of the cylinders:

- For YC3 (symmetric anisotropy), the $K$ points are on the allowed momentum cuts, so slightly changing $t'$ relative to $t$ would be expected to reduce the strength of the ordering. This prediction does not match our data. The most like reason is that, as seen in the phase diagram (Figure 2(c) of the main text), the small cylinder circumference itself produces an effective anisotropy, so that the point with effective $t'/t = 1$ seems to in fact be at more like $t'/t = 1.2$. In that case, we expect that $t'/t$ should be increased slightly in order to maximize the spin ordering at the $K$ points.

  The spiral order for $t' \gg t$ is probably unobservable given that there is no momentum cut near the expected location of that peak.

- For YC4 with symmetric anisotropy, the $K$ points are not on the allowed momentum cuts. Rather, the nearest momentum cuts correspond to the spiral order that appears at a slightly larger value of $q$ than $2\pi/3$, closer to the Néel point, so we expect that $t'/t$ slightly below 1 would give the maximum of spiral order. To be precise, the momentum cuts include the expected locations of the structure factor peaks when $q = 3\pi/4$, which is expected classically for $t'/t = 1/\sqrt{-2\cos(q)} \approx 0.84$. A weak spiral order could also appear on the $k_y = \pi$ momentum cut, between the upper and lower pairs of $M$ points.

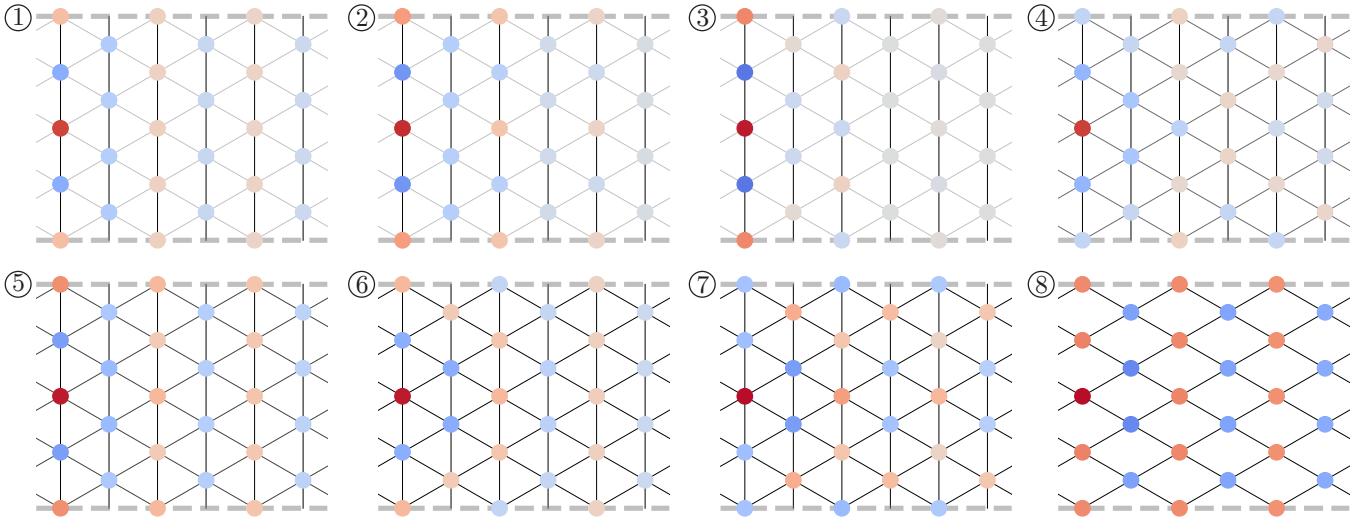

FIG. S7. Data for YC4 with symmetric anisotropy. Real-space correlations $\langle S_z S_z \rangle$ at a representative point in each phase, with labels corresponding to Figure 3(a) in the main text. The color of each point shows the correlation of $S_z$ on that site with $S_z$ on the dark red site at left. Red indicates positive correlations, while blue indicates negative correlations. The anisotropy in the phase is indicated by the strength of the lines showing the lattice.

- For YC4 with asymmetric anisotropy, the $K$ points are again not on the allowed momentum cuts, but due to the different orientation, the line of classically expected structure factor peaks now intersects the momentum cut that is closest to the $K$ points at $q = \pi/2$ (at $k_x = \pi\sqrt{3}/2$), when $t'/t \to \infty$. However, as noted above, linear spin wave theory suggests the spiral order goes to zero in this limit, so in practice we expect to observe a maximum at some finite $t'/t > 1$.

- For YC6 (symmetric anisotropy), the $K$ points are included on the allowed momentum cuts, so we expect the spiral order to be strongest exactly at the isotropic point. The large $t'/t$ limit of the spiral order, for which the peaks are on the line between $M$ points, is also allowed on this cylinder. Furthermore, there is an allowed momentum cut at $k_y = \pi/3$, which is expected to contain a spiral order peak for $q = 5\pi/6$, which classically would be at $t'/t \approx 0.76$. In our simulations, we find that this $t'/t$ is already within the Néel phase, but spiral order with this structure factor peak is possibly present at $t'/t = 0.9$.

## II. ADDITIONAL DATA

Here we present additional data from our simulations, beyond what is shown in Figures 3-6 of the main text.

### A. YC4 symmetric

#### 1. Data in each phase

We first show some additional quantities for each phase: the real-space $\langle S_z S_z \rangle$ correlations (Figure S7) and the spin- and momentum-resolved entanglement spectrum (Figure S8). In each figure, the phases are labeled with the numbers from Figure 3(a) of the main text.

#### 2. Open vs closed Fermi surface

One characteristic of the 1D metal phase, labeled by ① in Figure 3(a) of the main text, is that the $k$-space occupation numbers indicate an open Fermi surface, similar to the $U = 0$ state for $t'/t \gtrsim 1.636$. Here we first illustrate the transition from open to closed Fermi surface (interpreted loosely, since the lower phase, labeled ④, is likely a Luther-Emery liquid and not actually a Fermi liquid) by showing the computed $k$-space occupations at

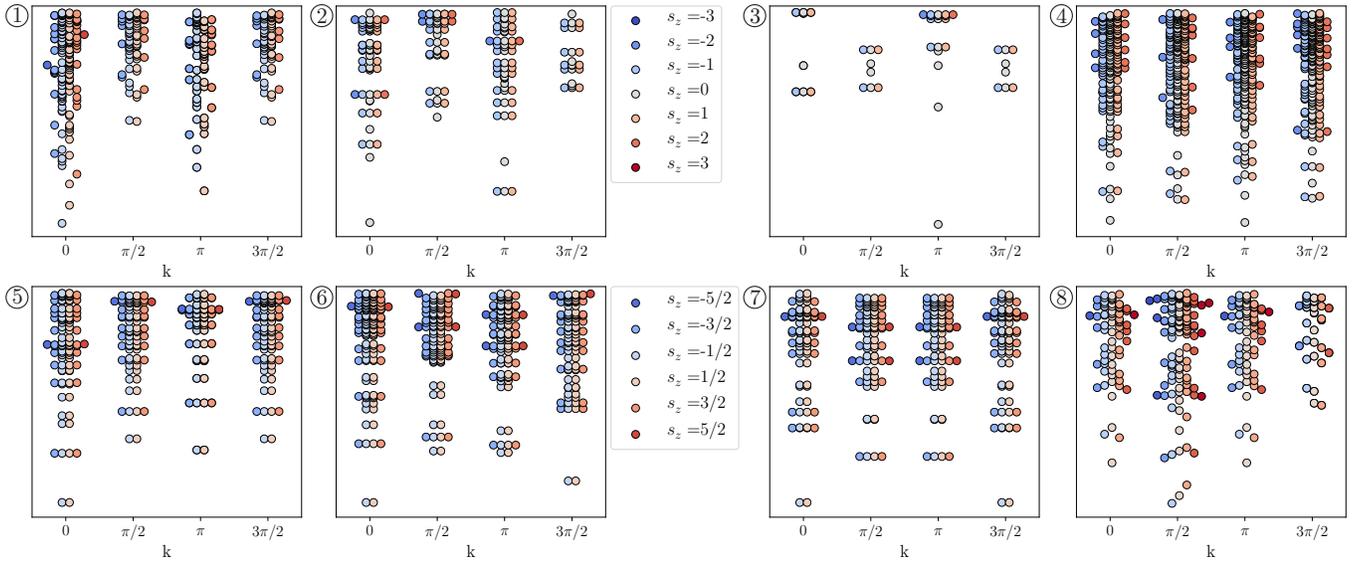

FIG. S8. Data for YC4 with symmetric anisotropy. Spin- and momentum-resolved entanglement spectrum at a representative point in each phase, with labels corresponding to Figure 3(a) in the main text. The vertical scale is the same in each figure. We show only Schmidt values with charge quantum number 0. Colors indicate spin quantum numbers, according to the keys on panels ② and ⑥; Schmidt values are shown with a horizontal offset proportional to the spin for clarity. Note that in most phases the Schmidt values are nicely organized into spin multiplets, the exceptions being the low-$U$ phase, where the wavefunctions are less well converged at this bond dimension, and for the Néel phase where translation symmetry of $\langle S_z \rangle$ is broken.

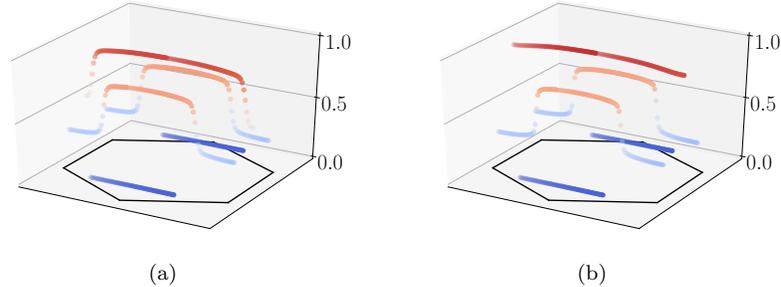

|              (a)              |              (b)              |

FIG. S9. Data for YC4 with symmetric anisotropy. Occupation on allowed momentum cuts for $U/t = 6$, (a) $t'/t = 1.6$ (b) $t'/t = 1.7$. The occupation numbers reveal what appears to be the opening of the Fermi surface, which at $U = 0$ happens around $t'/t \approx 1.636$.

$U/t = 6$ and $t'/t = 1.6$ and $1.7$, on either side of the transition; see Figure S9(a) and (b). Evidently, the opening of the Fermi surface can be detected by the occupation at the edge of the Brillouin zone on the $k_y = 0$ line, so we plot this quantity for every point of the phase diagram in Figure S10.

### 3. Alternating charge current phase

Recall that this phase has a scalar chiral order parameter whose sign alternates with a two ring unit cell, and has charge currents running around the cylinder in opposite directions on neighboring rings. In Figure S11 we illustrate the pattern of scalar chirality and local charge currents that we observe. For the charge currents, recall that the current on the bond between sites 1 and 2 is

$$J \propto i \sum_\sigma c_{1\sigma}^\dagger c_{2\sigma} - c_{2\sigma}^\dagger c_{1\sigma} \tag{14}$$

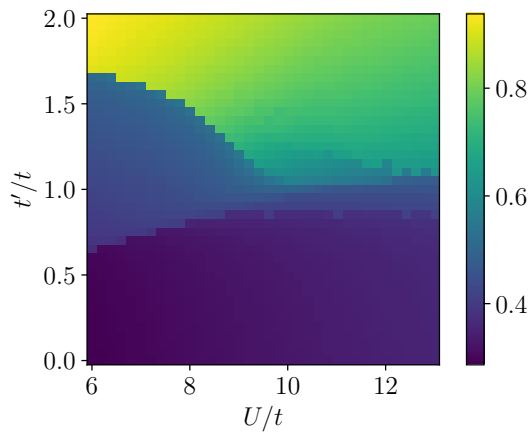

FIG. S10. Data for YC4 with symmetric anisotropy. Occupation at the edge of the Brillouin zone with $k_y = 0$. The boundary between the metal/Luther-Emergy liquid phase and the 1D metal phase is distinct.

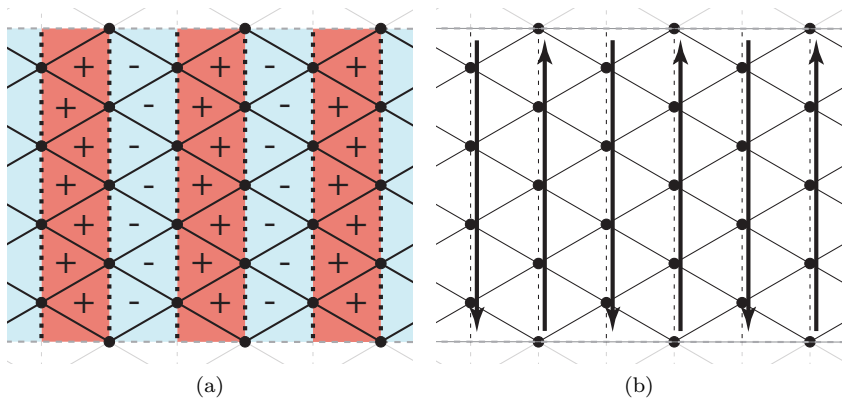

(a)  (b)

FIG. S11. Data for YC4 with symmetric anisotropy. (a) Observed sign pattern of scalar chiral order parameter in alternating charge current phase. (b) Observed pattern of local charge currents.

and consequently we measure the current by the imaginary part of $\sum_\sigma \langle c_{1\sigma}^\dagger c_{2\sigma} \rangle$. (The spin current would be the same, but the contributions for the two spins are subtracted rather than added. We find no net spin current.)

### 4. Gapless spin liquid phase

We showed in the main text that this phase has a very large correlation length for charge 0 excitations, meaning that it is likely to be spin gapless. By performing spin flux insertion, we scan the allowed momentum cuts through the Brillouin zone, revealing the locations of gapless points in momentum space. For details about this technique, see the Supplemental Material of our previous paper [5].

The transfer matrix spectrum is shown as a function of flux and as a function of momentum $k_y$ in Figure S12. Evidently, the correlation length is large (corresponding to small values in the transfer matrix spectrum) and hence the spin gap is small only with flux near 0 and $4\pi$, corresponding to small momentum. Accordingly, the data suggest that there either is a small gapless region, or possibly that at higher bond dimension there would be only isolated gapless points at 0 and $4\pi$ flux, corresponding to $k_y = 0$, consistent with a Dirac spin liquid (DSL).

Note that while the figures include eigenvalues for charge excitations, the low-lying points all have charge 0. In other words, charge excitations are gapped.

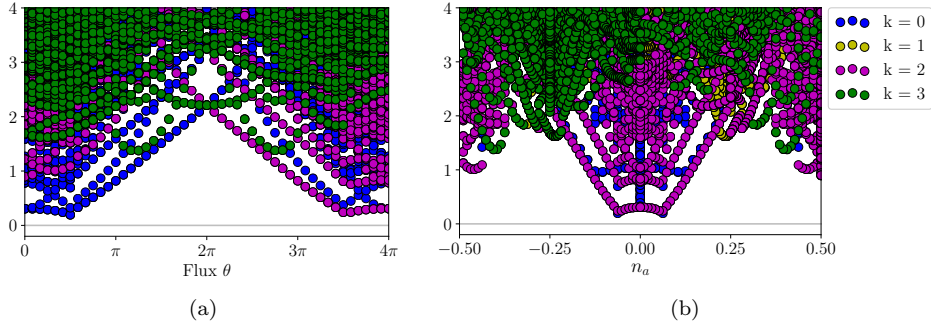

(a)                                                    (b)

FIG. S12. Data for YC4 with symmetric anisotropy. Transfer matrix spectrum in the gapless spin liquid phase, at $U/t = 10$, $t'/t = 1.15$. Color indicates quantum number for momentum around the cylinder. (a) As a function of inserted spin flux. (b) As a function of momentum $k_y$. To be precise, the horizontal axis is $n_a$, giving momentum as a fraction of the reciprocal lattice vector $\mathbf{k}_a$. (There is also a component along $\mathbf{k}_b$, but that vector has no $y$ component so is not needed to determine $k_y$.)

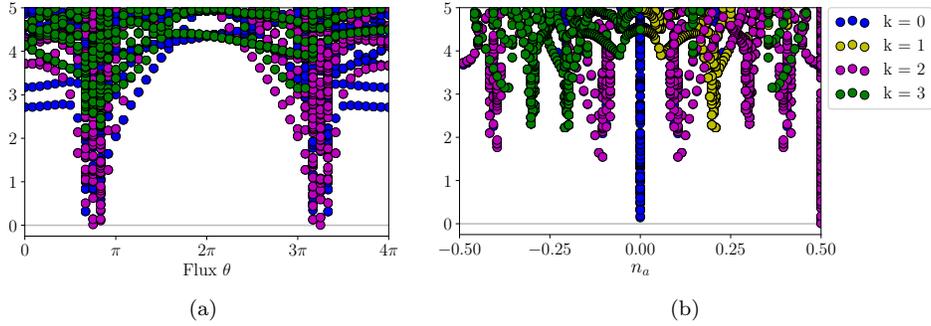

(a)                                                    (b)

FIG. S13. Data for YC4 with symmetric anisotropy. Transfer matrix spectrum in the 1D spin liquid phase, at $U/t = 13$, $t'/t = 2$. Color indicates quantum number for momentum around the cylinder. (a) As a function of inserted spin flux. (b) As a function of momentum $k_y$, given in terms of $n_a$, a fraction of the reciprocal lattice vector $\mathbf{k}_a$.

### 5. High-$U$ flux insertion

We have also performed spin flux insertion at $U/t = 13$ for the full range of anisotropy from $t'/t = 0$ to 2, with two primary goals: (1) understand the upper right phase, and (2) check whether the phase boundaries shift significantly with flux insertion.

We first show the transfer matrix spectrum, again plotted against both flux and $k_y$, for $U/t = 13$ and $t'/t = 2$ in Figure S13. There are again isolated gapless points, but at very different locations than for the gapless spin liquid phase discussed above. These gapless points are plausibly consistent with the 1D spin liquid found using variational Monte Carlo [6].

For comparison, we also show the transfer matrix spectrum as a function of flux in the other two observed high-$U$ phases, in Figures S14 and S15. For the spiral phase we use $t'/t = 0.95$, as this is the only value of anisotropy for which the spin structure factor remains qualitatively the same at all values of flux. For $t'/t \geq 1$, near $2\pi$ flux the structure factor looks like that of the 1D spin liquid, while for $t'/t \leq 0.9$, near $2\pi$ flux the structure factor looks like that of the Néel phase.

To check the general behavior of the three high-$U$ phases with flux insertion, we first plot correlation lengths in several charge sectors as a function of $t'/t$ and flux in Figure S16. The lines of maxima in all three charge sectors for $t'/t > 1$ correspond to the gap closing points in the transfer matrix spectra of Fig. S13.

In Figure S17 we similarly show the height of the spin structure factor at various points in the Brillouin zone. Panel (a) shows the $\langle S_z S_z \rangle$ structure factor, which demonstrates no significant shift in phase boundaries, though as mentioned above, the spiral phase appears smaller near $2\pi$ flux.

Panels (b) and (c) show the $\langle S_- S_+ \rangle$ structure factor, focusing particularly on the spiral order. As discussed in Section I C above, spin wave analysis of the Heisenberg model on the anisotropic triangular lattice suggests that the most stable spiral order will be incommensurate on the YC4 cylinder for most $t'/t$ [4]. This is equivalent to the

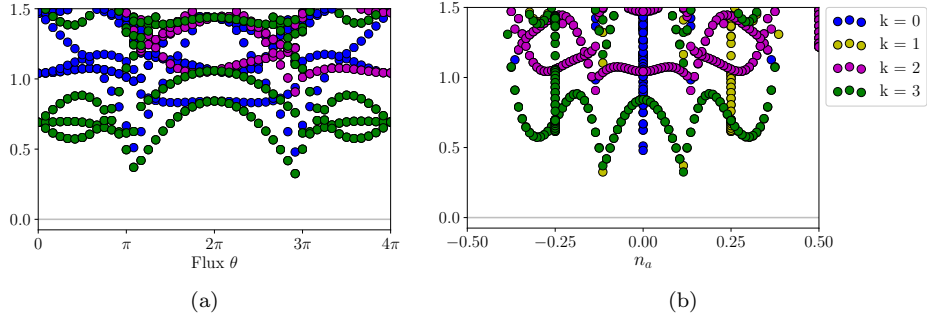

FIG. S14. Data for YC4 with symmetric anisotropy. Transfer matrix spectrum in the spiral phase, at $U/t = 13$, $t'/t = 0.95$. Color indicates quantum number for momentum around the cylinder. (a) As a function of inserted spin flux. (b) As a function of momentum $k_y$, given in terms of $n_a$, a fraction of the reciprocal lattice vector $\mathbf{k}_a$.

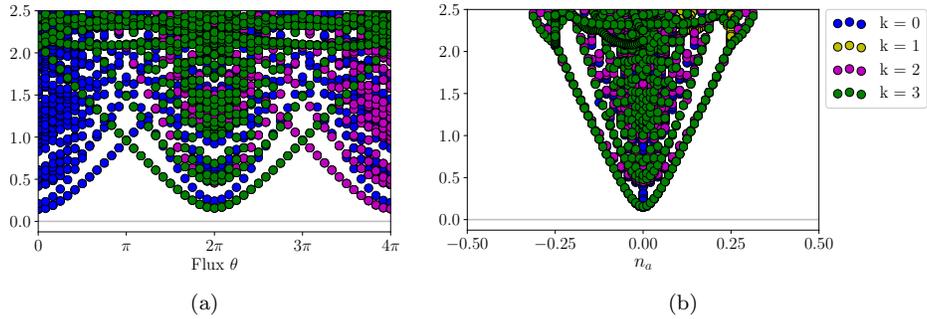

FIG. S15. Data for YC4 with symmetric anisotropy. Transfer matrix spectrum in the Néel phase, at $U/t = 13$, $t'/t = 0$. Color indicates quantum number for momentum around the cylinder. (a) As a function of inserted spin flux. (b) As a function of momentum $k_y$, given in terms of $n_a$, a fraction of the reciprocal lattice vector $\mathbf{k}_a$.

statement that the expected peak of the spin structure factor will not be on the allowed momentum cuts; however, the allowed momentum cuts for $\langle S_- S_+ \rangle$ shift with flux insertion, so that more general spiral orders can become effectively commensurate for these spin components.

In (b) we show the $\langle S_- S_+ \rangle$ structure factor at a number of values of inserted flux for $t'/t = 1$, and we indicate the momentum cut that can contain the spiral order peak for general $t'/t$. In (c), we show versus anisotropy and flux the difference between the maximum and minimum height on this cut, an indication of whether there is an isolated peak corresponding to spiral order. The dashed black line indicates the flux for which the allowed momentum cut would contain the spiral order peak in the spin wave analysis for each value of $t'/t \geq 1$.

The observed behavior at the isotropic point $t'/t = 1$ validates the picture that shifting momentum cuts with flux insertion can lead to stabilized magnetic order when the cuts contain the expected structure factor peak from the two-dimensional model: the $\langle S_- S_+ \rangle$ peak height is maximized at exactly the flux where the momentum cuts include the $K$ points, the location of the peaks for the 120° magnetically-ordered phase. Moving away from the isotropic point, the spiral order does not suddenly vanish as in the $\langle S_z S_z \rangle$ structure factor in panel (a)ⓒ. Some signature of the spiral phase at first glance extends to large $t'/t$, but specifically along the dashed line indicating the classical/spin-wave spiral order, the tendency towards ordering seems to decay quickly. Overall, the data may suggest that the spiral order would be limited to a finite range of $t'/t$ even in two dimensions where commensurability is not an issue, though the data are by no means conclusive.

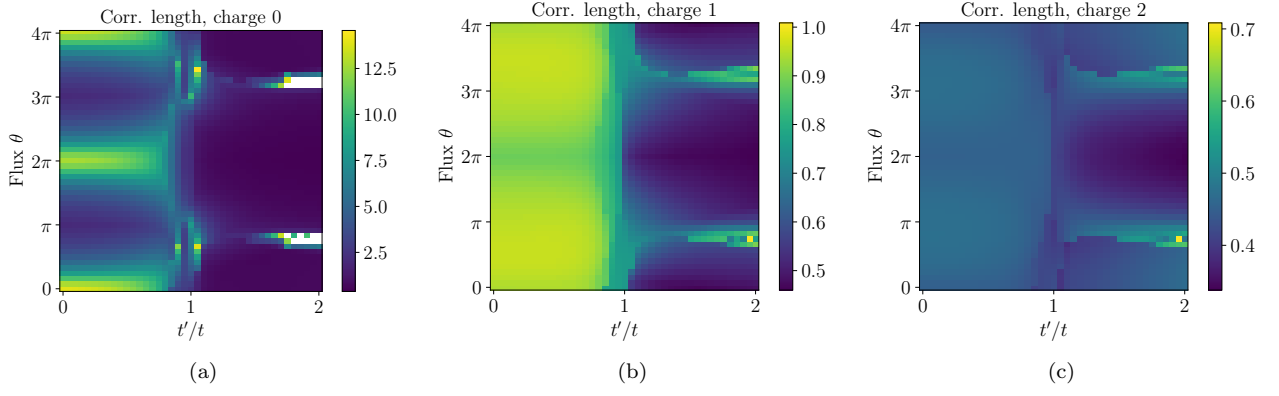

FIG. S16. Data for YC4 with symmetric anisotropy. Here we show correlation lengthsas as a function of anisotropy $t'/t$ and spin flux $\theta$ on the $U/t = 13$ line. (a) Correlation length for excitations with charge 0. White regions on the right indicate that the correlation length is larger than the cutoff value of 15. (b) Correlation length for excitations with charge 1. (c) Correlation length for excitations with charge 2.

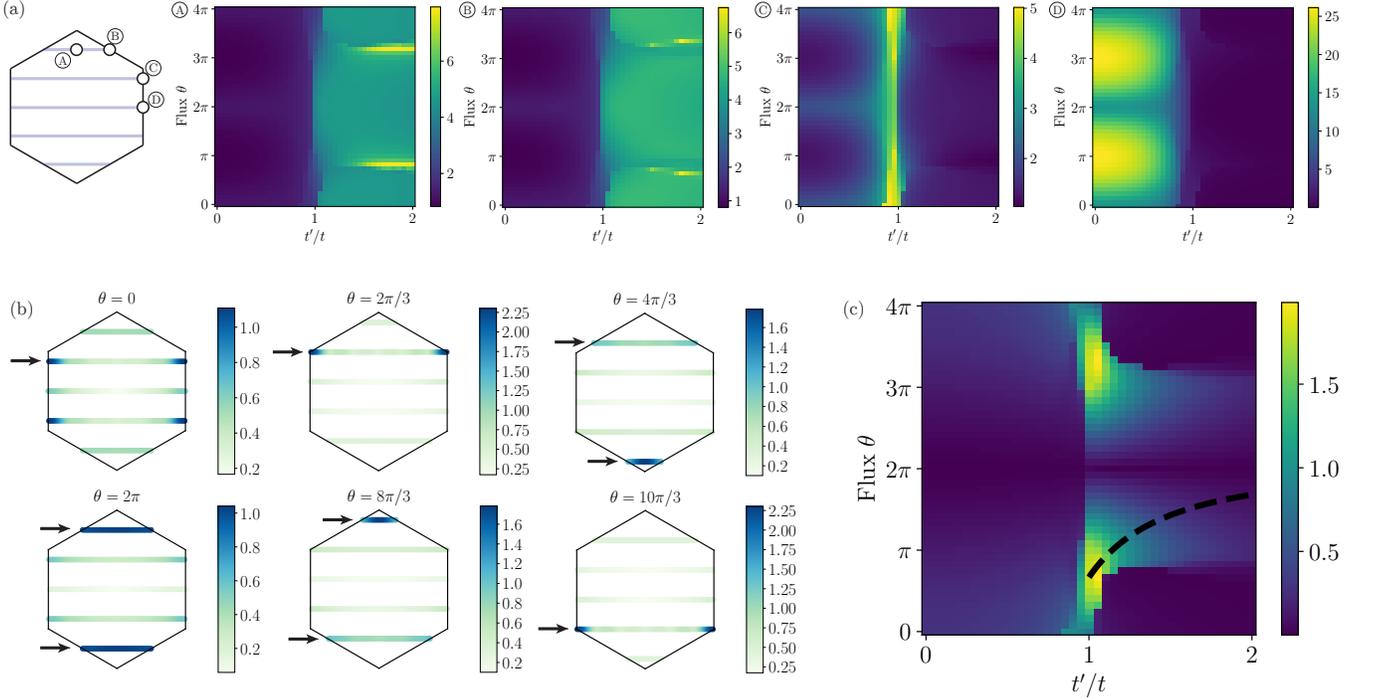

FIG. S17. Data for YC4 with symmetric anisotropy. Here we show the spin structure factor as a function of anisotropy $t'/t$ and spin flux $\theta$ on the $U/t = 13$ line. (a) Height of $\langle S_z S_z \rangle$ structure factor at points in the Brillouin zone labeled by the circled letters. See Figure 3 of the main text for a description of each of these points. (b) $\langle S_- S_+ \rangle$ structure factor for $t'/t = 1$ with inserted spin flux at multiples of $2\pi/3$. The allowed momentum cuts shift with flux insertion, so that the 120° order is effectively commensurate with $2\pi/3$ flux, while more general spiral orders for $t'/t > 1$ are commensurate for some flux between $2\pi/3$ and $2\pi$. (c) Considering the momentum cut indicated in the structure factors of (b), we show for each $t'/t$ and $\theta$ the difference between the maximum and minimum heights on this cut. For each $t'/t \geq 1$, for the value of flux indicated by the dashed black line this cut will contain the expected peak for spiral order from linear spin wave analysis [4]. If there is indeed spiral order, there will be an isolated peak and the height difference will be large. Otherwise, the difference will be small, either because there is no significant ordering on this momentum cut or because, as in the case of $2\pi$ flux in (b), the structure factor is uniform along the cut.

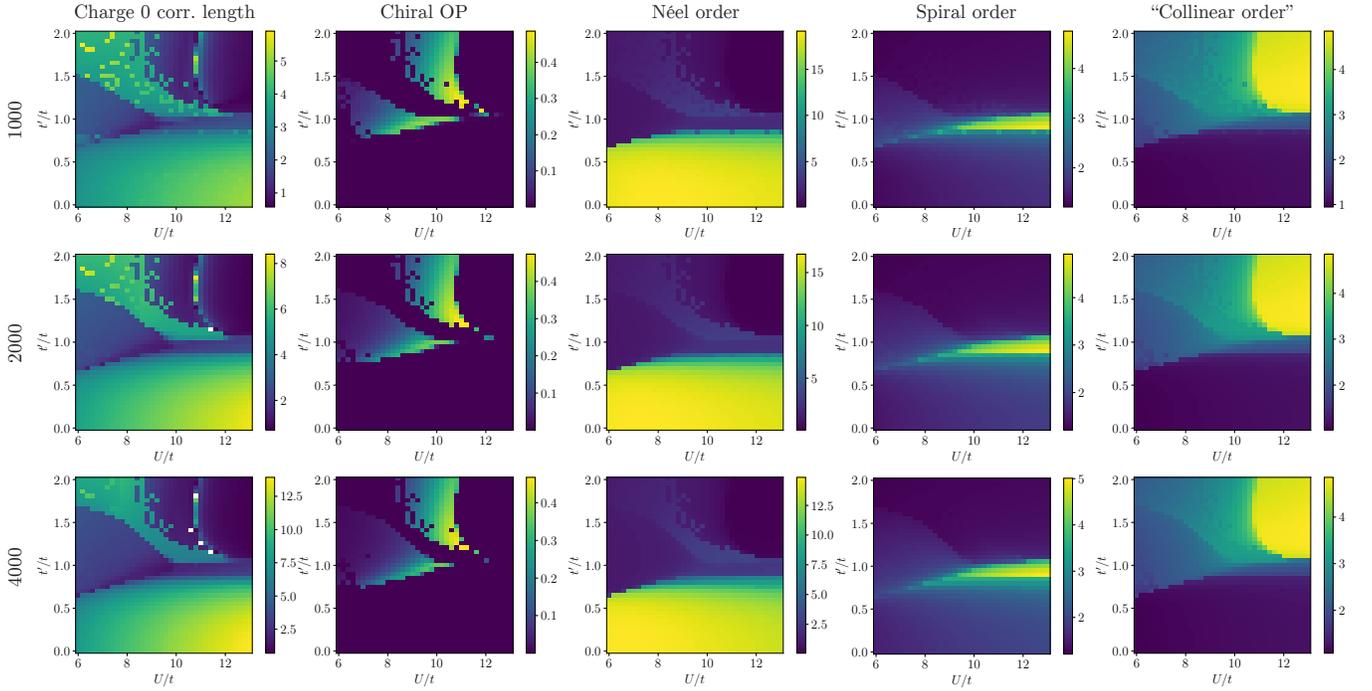

FIG. S18. Data for YC4 with symmetric anisotropy. Comparison of results for a variety of quantities, computed using bond dimensions of 1000, 2000, and 4000. The qualitative behavior is unchanged, apart from a shift in phase boundaries. Note that the third row contains precisely the figures shown as part of Figure 3 in the main text.

### 6. Smaller bond dimension data

Finally, we show some representative data with smaller bond dimensions of 1000 and 2000, compared with 4000 used in the main text figures and above. In Figure S18 we show side-by-side comparisons of the results for different bond dimensions for the following quantities: scalar chiral order parameter, charge 0 correlation length, and spin structure factor height at the right edge of the Brillouin zone with $k_y = 0$, $\pi/2$, and $\pi$. There is a significant shift in phase boundaries at low $U$, but otherwise the qualitative behavior is the same at each bond dimension. Note that, as stated in the main text, the plotted data results from taking for each point the lowest energy among several independent data sets; although some use lower bond dimension results to seed higher bond dimensions, others do not, so this is not likely to cause an artificial similarity between the results at different bond dimensions.

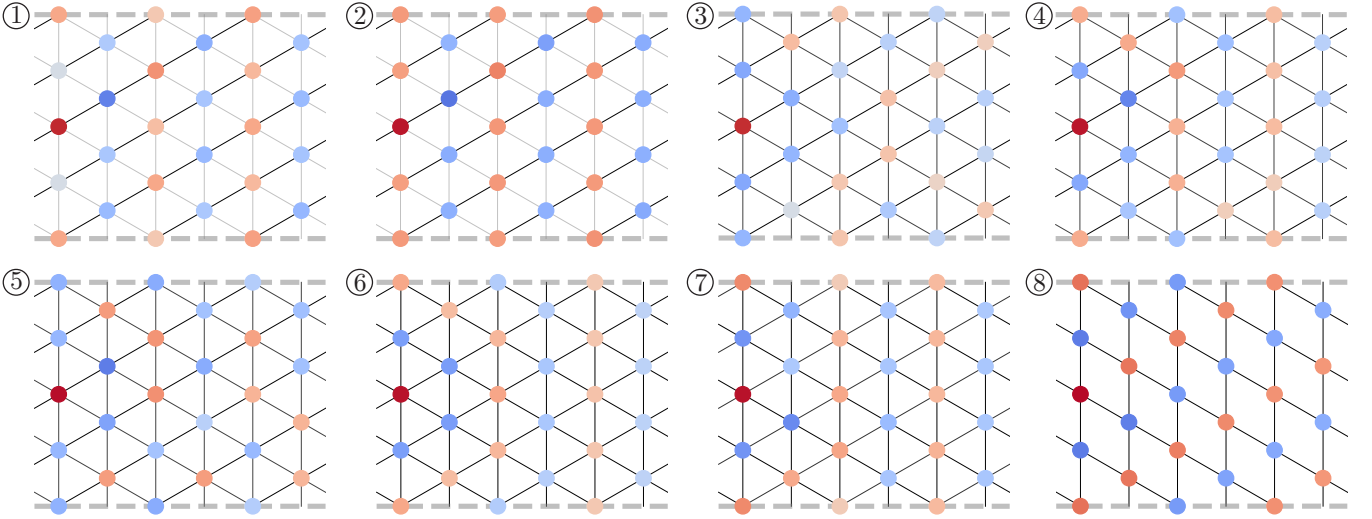

FIG. S19. Data for YC4 with asymmetric anisotropy. Real-space correlations $\langle S_z S_z \rangle$ at a representative point in each phase, with labels corresponding to Figure 4(a) in the main text. The color of each point shows the correlation of $S_z$ on that site with $S_z$ on the dark red site at left. Red indicates positive correlations, while blue indicates negative correlations. The anisotropy in the phase is indicated by the strength of the lines showing the lattice.

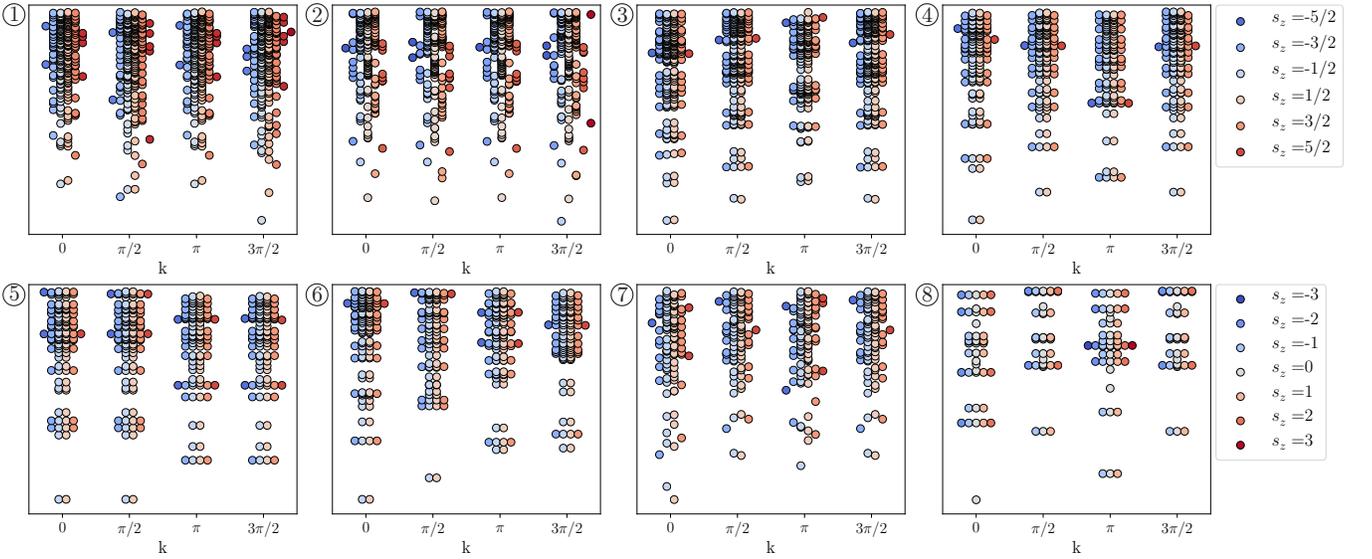

FIG. S20. Data for YC4 with asymmetric anisotropy. Spin- and momentum-resolved entanglement spectrum at a representative point in each phase, with labels corresponding to Figure 4(a) in the main text. The vertical scale is the same in each figure. We show only Schmidt values with charge quantum number 0. Colors indicate spin quantum numbers according to the keys on panels ④ and ⑧; Schmidt values are shown with a horizontal offset proportional to the spin for clarity. Here the phases in which the Schmidt values are not clearly organized into spin multiplets are the two upper phases, ① and ②, which break translation symmetry, the gapless spin liquid phase ⑦.

## B. YC4 asymmetric

### 1. Data in each phase

We first show some additional quantities for each phase: the real-space $\langle S_z S_z \rangle$ correlations (Figure S19) and the spin- and momentum-resolved entanglement spectrum (Figure S20). In each figure, the phases are labeled with the numbers from Figure 4(a) of the main text.

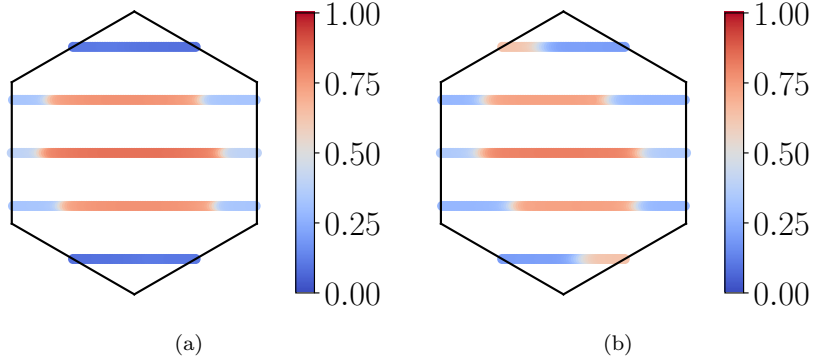

FIG. S21. Data for YC4 with asymmetric anisotropy. Occupation on allowed momentum cuts for $U/t = 6$, (a) $t'/t = 1.4$ (b) $t'/t = 2.0$. The occupation numbers reveal what appears to be the opening of the Fermi surface, which at $U = 0$ happens around $t'/t \approx 1.636$.

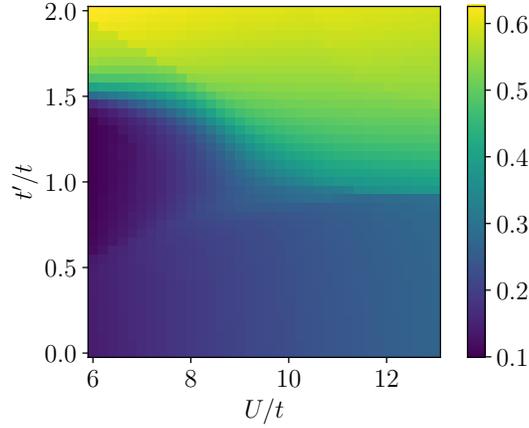

FIG. S22. Data for YC4 with asymmetric anisotropy. Occupation at the edge of the Brillouin zone with $k_y = 0$. The boundary between the metal/Luther-Emergy liquid phase and the 1D metal phase is not distinct.

### 2. Open vs closed Fermi surface

Compared with the YC4 cylinder with symmetric anisotropy, in this case it is less clear whether what appears to be the opening of the Fermi surface based on $k$-space occupation actually corresponds to a phase transition. Nevertheless, it is still instructive to consider the occupation as a signature of the various phases. We first show in Figure S21 the occupation on allowed momentum cuts for two points with $U/t = 6$ and different degrees of anisotropy, namely $t'/t = 1.4$ and 2.0. Since the underlying Fermi surface is effectively rotated compared with the symmetric anisotropy case, the "opening point" now lies at one of the diagonal $M$ points. We then plot the occupation at this momentum for all parameter points, in Figure S22.

### 3. Gapless spin liquid

As in the case of symmetric anisotropy, we observe a phase that appears to be a gapless spin liquid, here just below the CSL. To better understand this phase, we again perform flux insertion and compute the transfer matrix spectrum, which shows the momentum of gapless excitations.

The transfer matrix spectrum as a function of flux and as a function of momentum $k_y$ is shown in Figure S23. The behavior looks quite similar to the gapless spin liquid just above the CSL in the symmetric case: the correlation length is again large and hence the spin gap is small only with flux near 0 and $4\pi$, corresponding to small momentum. So this phase as well is likely a nodal gapless spin liquid or else has a small gapless region.

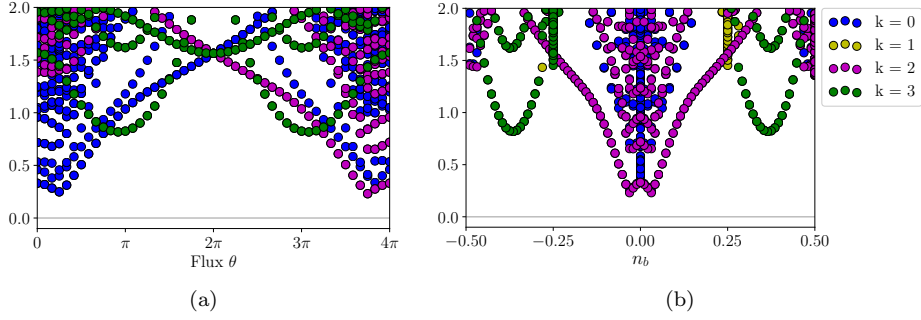

(a)                                    (b)

FIG. S23. Data for YC4 with asymmetric anisotropy. Transfer matrix spectrum in the gapless spin liquid phase, at $U/t = 10.4$, $t'/t = 0.9$. Color indicates quantum number for momentum around the cylinder. (a) As a function of inserted spin flux. (b) As a function of momentum $k_y$, measured in terms of $n_a$, a fraction of the reciprocal lattice vector $\mathbf{k}_a$.

The transfer matrix spectrum looks consistent with a Dirac spin liquid except right at zero flux. One explanation is that DMRG simply fails to converge to the gapless or nearly gapless states near 0 flux, a problem which could be solved except exactly at the gapless point by use of a higher bond dimension, in particular for the two data points between 0 flux and the DSL-like behavior; however, we tried this with bond dimensions of 8000 and 11314 ($\approx 8000\sqrt{2}$), compared with 4000 for the data in the figure, and did not find a significant difference. Possibly an even larger bond dimension would be needed.

Note that while the figures include eigenvalues for charge excitations, the low-lying points all have charge 0. In other words, charge excitations are gapped.

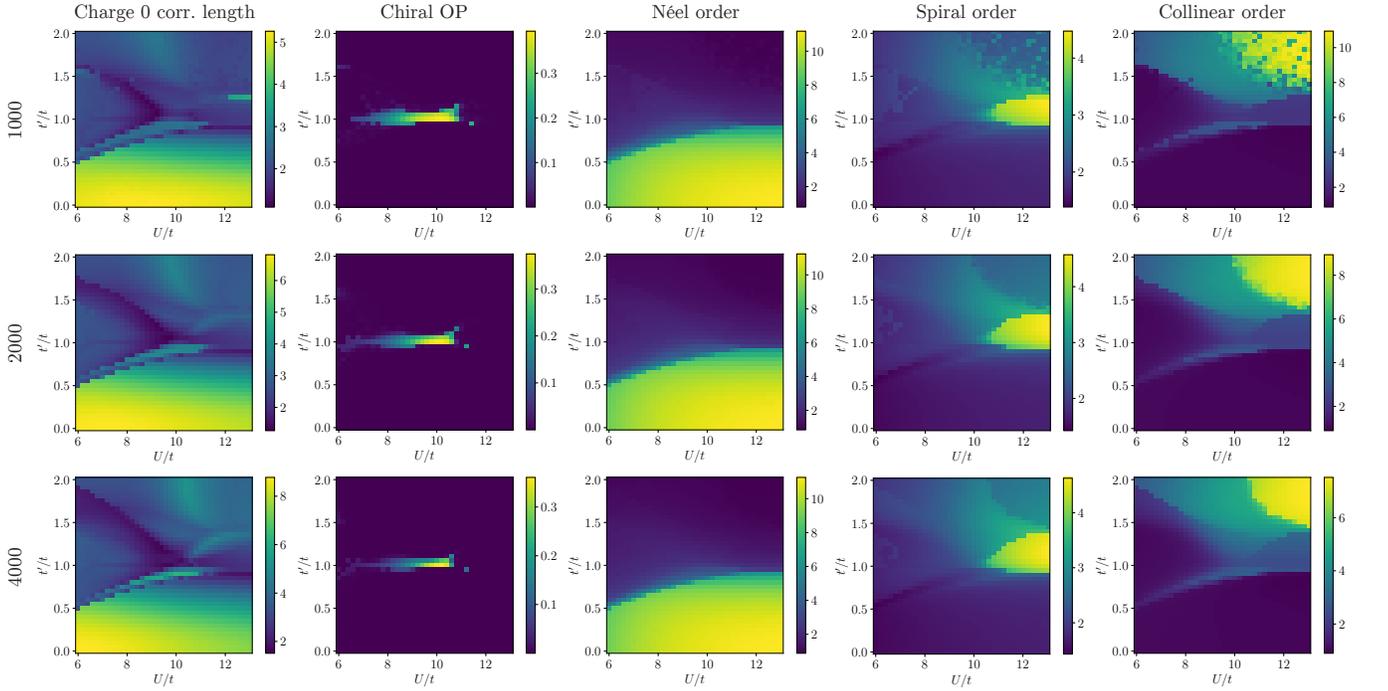

FIG. S24. Data for YC4 with asymmetric anisotropy. Comparison of results for a variety of quantities, computed using bond dimensions of 1000, 2000, and 4000. The main effect of increasing the bond dimension is apparently to shift the phase boundaries; the collinear order also seems less stable with small bond dimension. Note that the third row contains precisely the figures shown as part of Figure 4 in the main text.

#### 4. Smaller bond dimension data

Finally, we show some representative data with smaller bond dimensions of 1000 and 2000, compared with 4000 used in the main text figures and above. In Figure S24 we show side-by-side comparisons of the results for different bond dimensions for the following quantities: scalar chiral order parameter, charge 0 correlation length, spin structure factor height at $M$ points corresponding to Néel and collinear order, and max spin structure factor on the region marked by Ⓒ in Figure 4(b) of the main text.

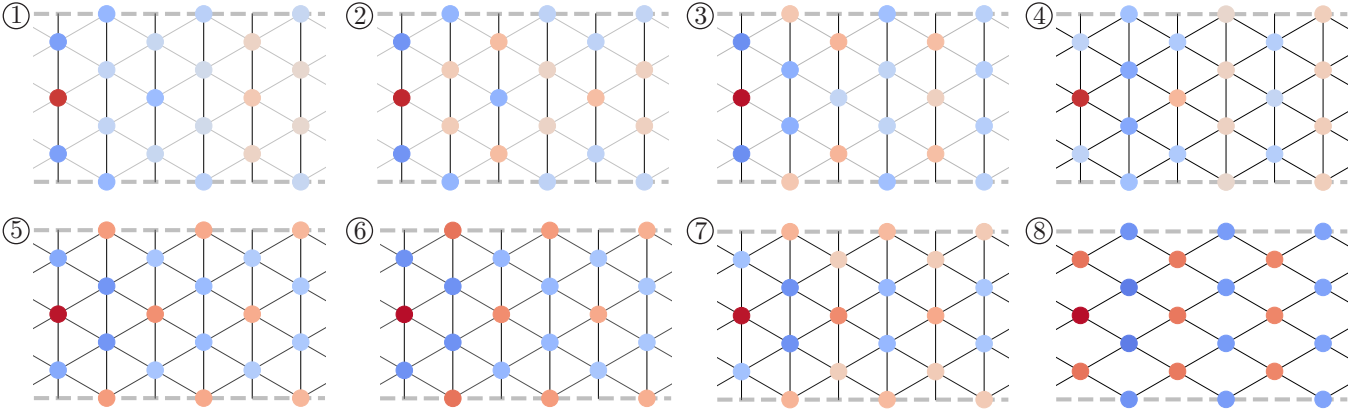

FIG. S25. Data for YC3 with symmetric anisotropy. Real-space correlations $\langle S_z S_z \rangle$ at a representative point in each phase, with labels corresponding to Figure 5(a) in the main text. The color of each point shows the correlation of $S_z$ on that site with $S_z$ on the dark red site at left. Red indicates positive correlations, while blue indicates negative correlations. The anisotropy in the phase is indicated by the strength of the lines showing the lattice.

### C.  YC3 symmetric

#### 1.  Data in each phase

We first show some additional quantities for each phase: the real-space $\langle S_z S_z \rangle$ correlations (Figure S25) and the spin- and momentum-resolved entanglement spectrum (Figure S26). In each figure, the phases are labeled with the numbers from Figure 5(a) of the main text.

#### 2.  Open vs closed Fermi surface

As with YC4 with symmetric anisotropy, the transition between the metal/Luther-Emery liquid phase and the 1D metal phase is clearly shown by the opening of the Fermi surface as measured by occupation number in momentum space. We show the occupation on allowed momentum cuts for one point on either side of the transition in Figure S27 and the occupation at the edge of the Brillouin zone on the $k_y = 0$ line for all points in the phase diagram in Figure S28.

#### 3.  Additional correlation length data

In the Figure 5 of the main text, parts (d) and (e), we show the correlation length for charge 0 excitations and for charge 1 excitations at each point in the phase diagram. Here, in Figure S29 we also show the correlation length for charge 2, and we additionally show the charge 0 data again but with all correlation lengths above 12 cut off, in order to get a better picture of the behavior in phases with shorter correlation lengths.

#### 4.  Chiral phases

Here we show three additional pieces of data. First, in Figure 5(c) of the main text we show the maximum magnitude of the scalar chiral order parameter over all plaquettes in the four ring unit cell. Here, in Figure S30 we show the minimum magnitude. There is only a small difference in the CSL phase, around the isotropic line, but a larger difference in the alternating chiral phase, ② in Figure 5(a) of the main text.

Next, we show the pattern of alternating sign of the scalar chiral order parameter in the CSL, as well as the pattern of local currents, in Figure S31. Note that the patterns are the same as for the alternating charge current phase of the YC4 symmetric cylinder. One difference, which is not shown in the figure, is that in the YC4 case the positive and negative scalar chiralities on different plaquettes are equal in magnitude, whereas for the YC3 CSL they are not—at bond dimension 4000, one is around 50% larger than the other, so that there is a net chirality.

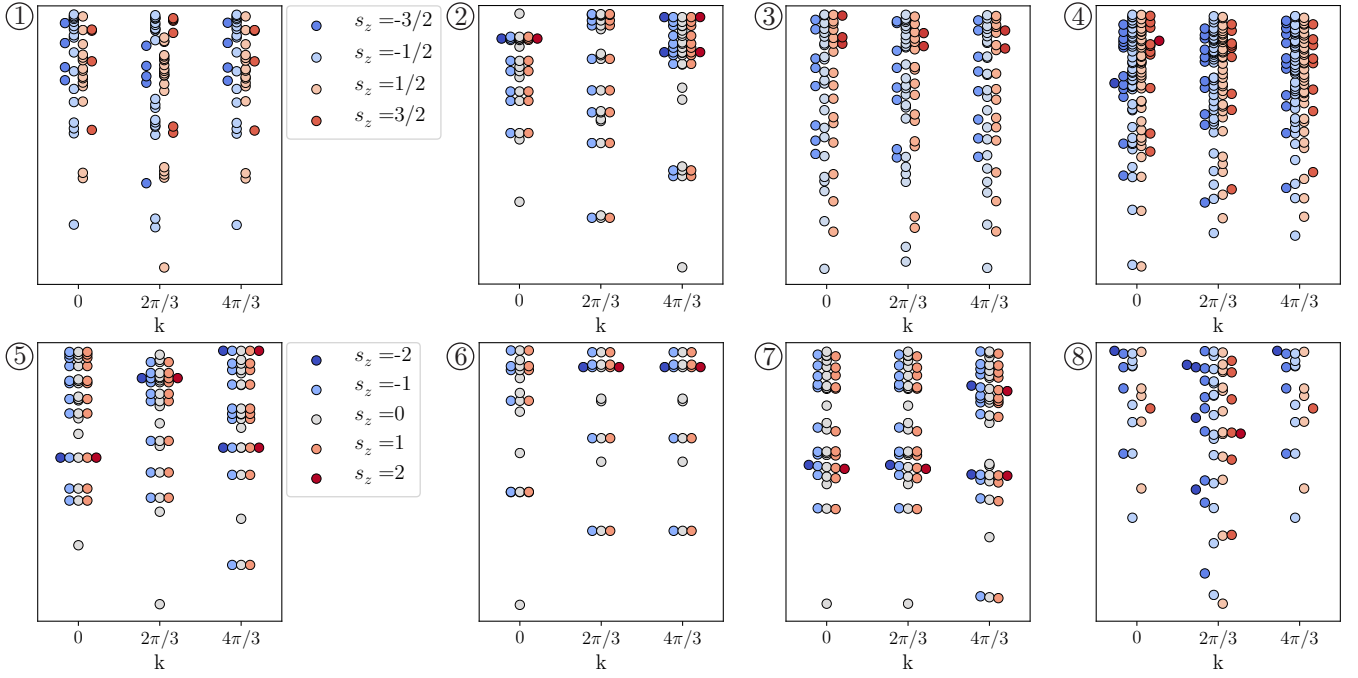

FIG. S26. Data for YC3 with symmetric anisotropy. Spin- and momentum-resolved entanglement spectrum at a representative point in each phase, with labels corresponding to Figure 5(a) in the main text. The vertical scale is the same in each figure. We show only Schmidt values with charge quantum number 0. Colors indicate spin quantum numbers according to the keys on panels ① and ⑤; Schmidt values are shown with a horizontal offset proportional to the spin for clarity. Note that in most phases the Schmidt values are nicely organized into spin multiplets, the exceptions being the low-$U$ phase, where the wavefunctions are less well converged at this bond dimension, and the Néel phase where translation symmetry of $\langle S_z \rangle$ is broken. Also note that the alternating chiral phase, labeled ②, has quite different entanglement spectra on cuts between different rings in the unit cell—for two of the cuts, time-reversal symmetry breaking is not evident, and on the other two it is; we have selected one of the latter.

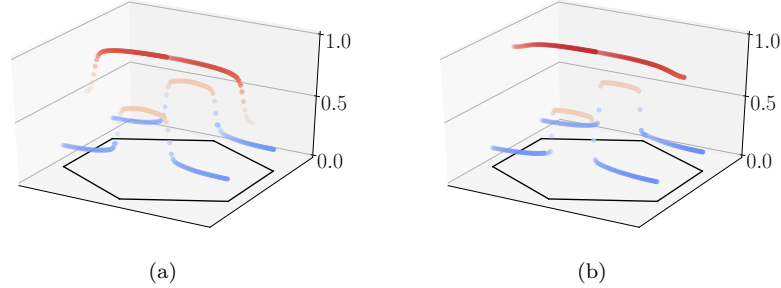

(a)                                      (b)

FIG. S27. Data for YC3 with symmetric anisotropy. Occupation on allowed momentum cuts for $U/t = 6$, (a) $t'/t = 1.6$ (b) $t'/t = 1.75$. The occupation numbers reveal what appears to be the opening of the Fermi surface, which at $U = 0$ happens around $t'/t \approx 1.636$.

Finally, we show the same patterns for the alternating chiral phase, labeled by ② in Figure 5(a) of the main text, in Figure S32. Here there is a two ring unit cell for the magnitude of the chiral order parameter, but a four ring unit cell when including the sign. The local charge currents also repeat with a four ring unit cell.

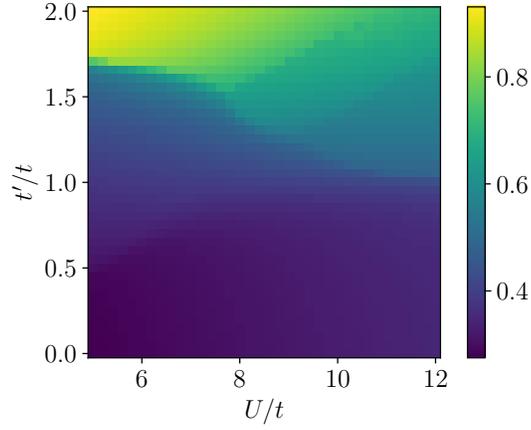

FIG. S28. Data for YC3 with symmetric anisotropy. Occupation at the edge of the Brillouin zone with $k_y = 0$. The boundary between the metal/Luther-Emergy liquid phase and the 1D metal phase is distinct.

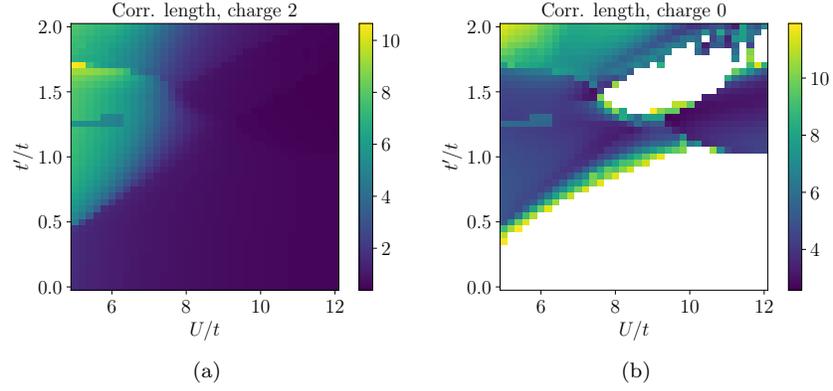

FIG. S29. Data for YC3 with symmetric anisotropy. (a) Correlation length for charge 2 excitations. (b) Correlations length for charge 0 excitations, only for points where the correlation length is below 12.

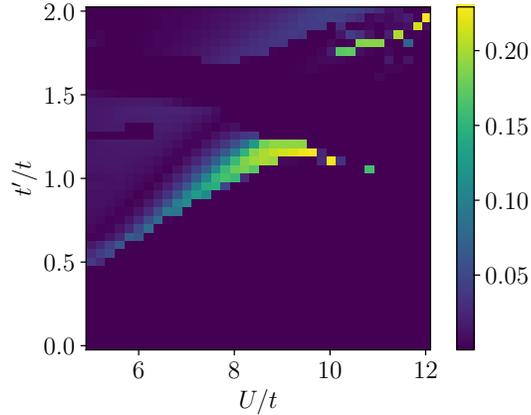

FIG. S30. Data for YC3 with symmetric anisotropy. For each parameter point, minimum magnitude of the scalar chiral order parameter over all plaquettes in the four ring unit cell. For comparison, see the maximum magnitude in Figure 5(c) of the main text.

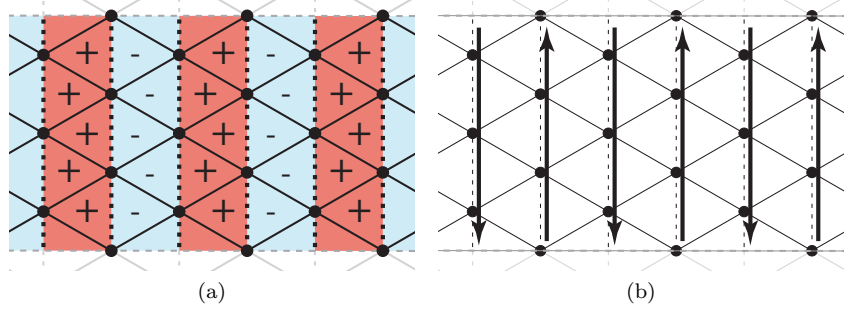

FIG. S31. Data for YC3 with symmetric anisotropy. (a) Sign pattern of scalar chiral order parameter in the CSL. (b) Pattern of local currents in the same phase.

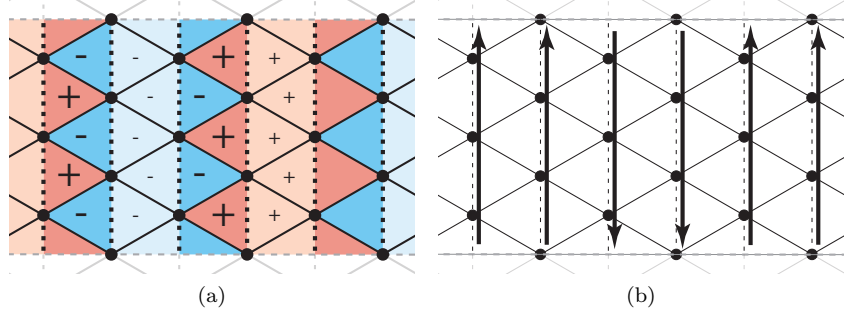

FIG. S32. Data for YC3 with symmetric anisotropy. (a) Sign pattern of scalar chiral order parameter in the alternating chiral phase. The darker colors/larger symbols indicate a larger magnitude. The larger magnitude is about an order of magnitude larger than the smaller one. (b) Pattern of local currents in the same phase.

### 5. Flux insertion for spin liquid

The final piece of additional data for the YC3 cylinder is the transfer matrix spectrum with flux insertion, shown as a function of flux in Figure S33. The phase appears gapless for most values of flux, apart from a gap just above zero flux. Note that unlike the transfer matrix figures above, we do not here plot versus momentum transfer because there appear to be gapless excitations at almost all momenta so the figure is not informative.

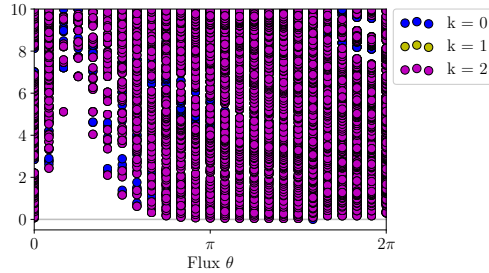

FIG. S33. Data for YC3 with symmetric anisotropy. Transfer matrix spectrum for spin liquid phase plotted vs flux inserted. Color indicates quantum number for momentum around the cylinder.

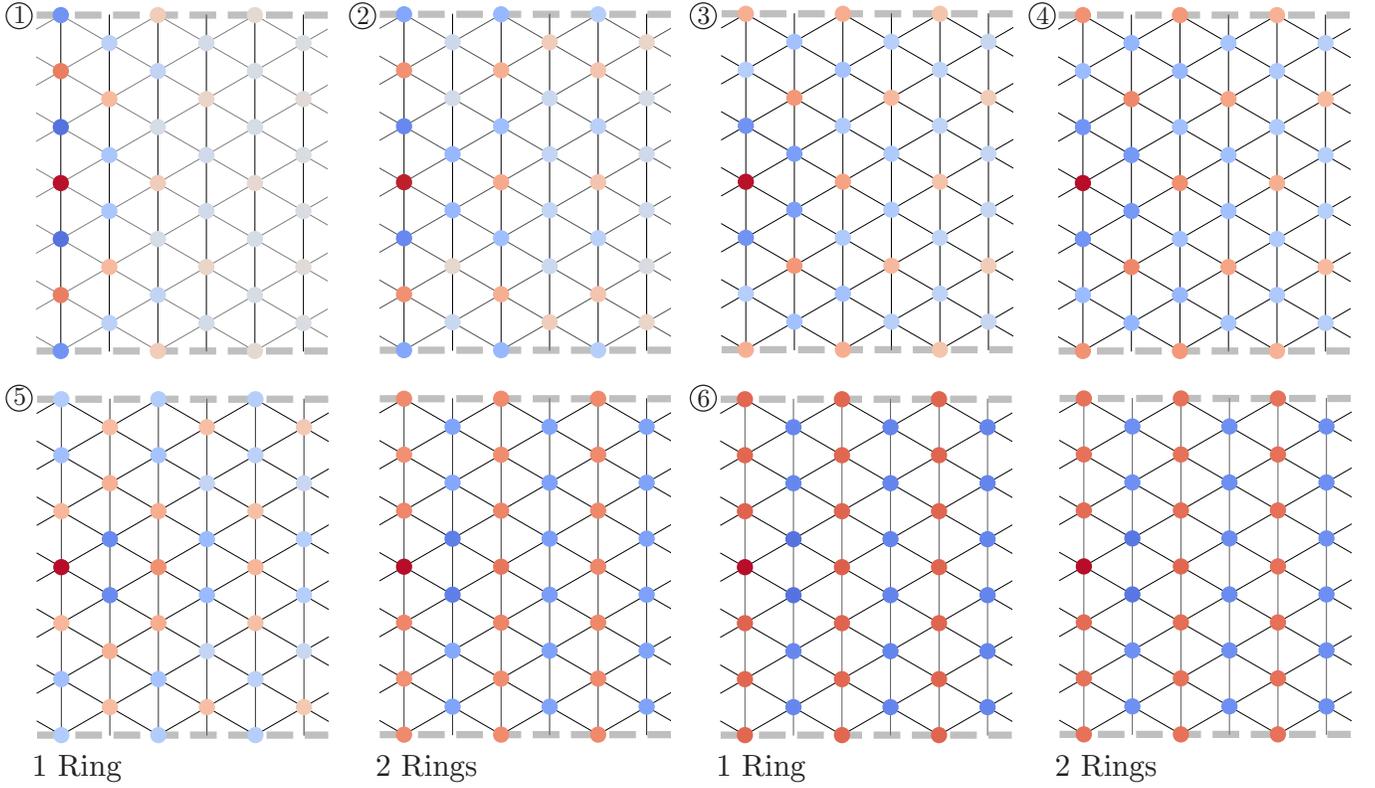

FIG. S34. Data for YC6 with symmetric anisotropy. Real-space correlations $\langle S_z S_z \rangle$ at a representative point in each phase, with labels corresponding to Figure 6(a) in the main text. The color of each point shows the correlation of $S_z$ on that site with $S_z$ on the dark red site at left. Red indicates positive correlations, while blue indicates negative correlations. The anisotropy in the phase is indicated by the strength of the lines showing the lattice. For the region labeled by ⑤, with a two ring unit cell the correlations are clearly the square lattice Néel order, while with one ring they look quite different. In contrast, in region ⑥, the Néel order is evident in both cases.

### D. YC6 symmetric

#### 1. Data in each phase

Here we show some additional quantities for each phase: the real-space $\langle S_z S_z \rangle$ correlations (Figure S34) and the spin- and momentum-resolved entanglement spectrum (Figure S35). In each figure, the phases are labeled with the numbers from Figure 6(a) of the main text. As in that figure, for the phases labeled ⑤ and ⑥ we show data from the ground state computed both with a one ring unit cell and with a two ring unit cell.

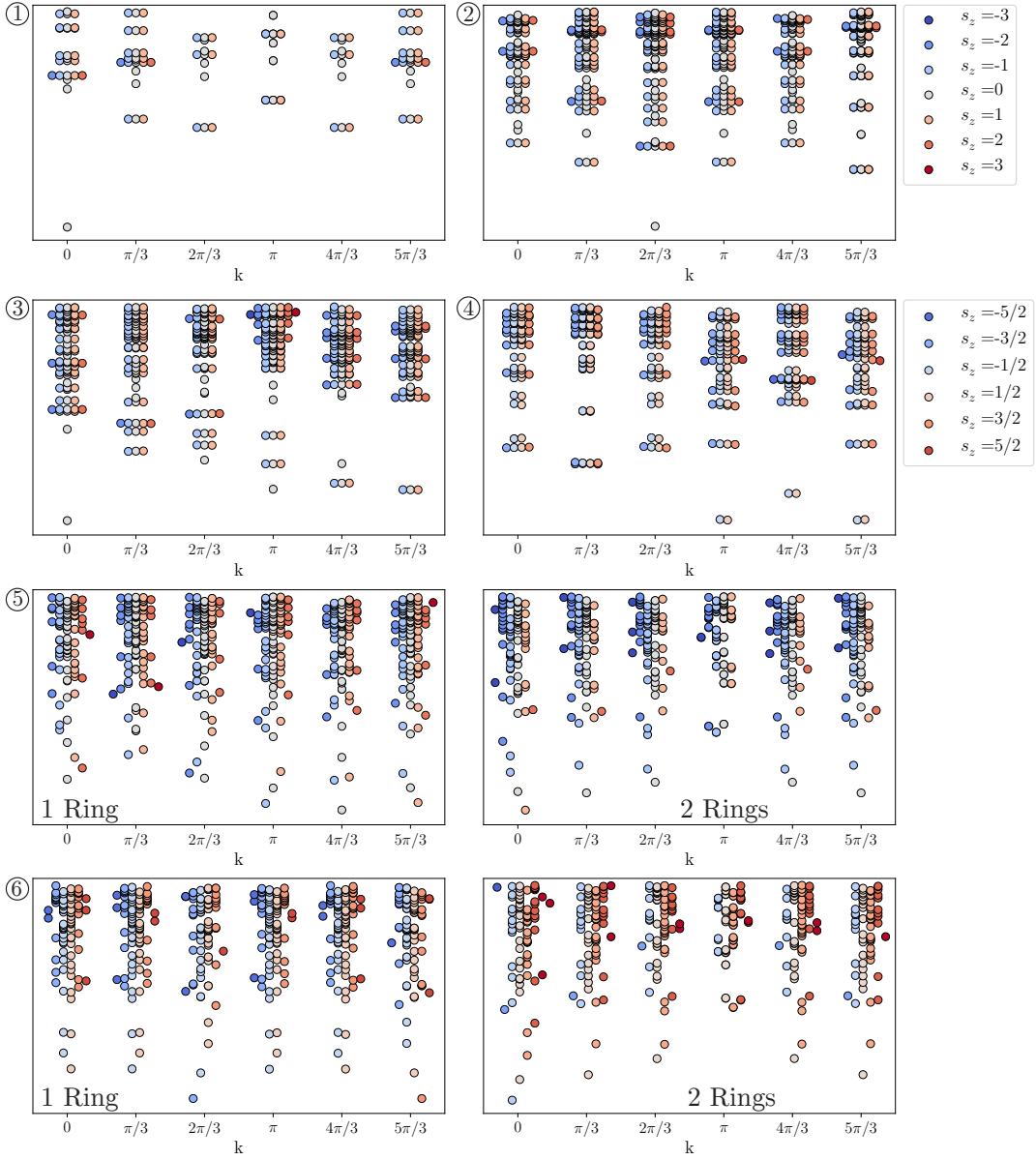

FIG. S35. Data for YC6 with symmetric anisotropy. Spin- and momentum-resolved entanglement spectrum at a representative point in each phase, with labels corresponding to Figure 6(a) in the main text. The vertical scale is the same in each figure. We show only Schmidt values with charge quantum number 0. Colors indicate spin quantum numbers according to the keys on panels ② and ④; Schmidt values are shown with a horizontal offset proportional to the spin for clarity.



\end{document}